\documentclass[pre,twocolumn,floatfix,amsmath,amssymb,showpacs]{revtex4}
\usepackage{dcolumn}
\usepackage{color}
\include{epsf}

\begin{document}
\title{On the emergence of very long time fluctuations and $1/f$
noise in ideal flows}

\author{
P. Dmitruk$^1$, P.D. Mininni$^{1,2}$, A. Pouquet$^2$, S. Servidio$^3$, 
and W.H. Matthaeus$^4$}

\affiliation{
$^1$ Departamento de F\'isica, Facultad de Ciencias Exactas y Naturales,
Universidad de Buenos Aires and \\
Instituto de F\'isica de Buenos Aires, CONICET, Argentina \\
$^2$ National Center for Atmospheric Research, Boulder, Colorado \\
$^3$ Dipartimento di Fisica, Universit\'a della Calabria, I-87036 Cosenza, Italy\\
$^4$ Bartol Research Institute and Department of Physics and Astronomy, 
University of Delaware, Newark, Delaware}

\begin{abstract}
This study shows the connection between three previously observed but
seemingly unrelated phenomena in hydrodynamic (HD) and 
magnetohydrodynamic (MHD) turbulent flows, involving the
emergence of  fluctuations occurring on very long time scales: the low-frequency $1/f$ noise 
in the power frequency spectrum, the delayed ergodicity
of complex valued amplitude fluctuations in wavenumber space, 
and the spontaneous flippings or reversals of large scale
fields. 
Direct numerical simulations of 
ideal MHD and HD 
are employed
in three space dimensions, at low resolution, for long periods 
of time, and with high accuracy to study
several cases:
Different geometries, presence of rotation and/or a uniform 
magnetic field, and different values of the associated conserved global
quantities. 
It is conjectured that the 
origin of all these long-time phenomena is 
rooted in the interaction 
of the longest wavelength 
fluctuations 
available to the system with fluctuations 
at much smaller scales. 
The strength of this non-local interaction 
is controlled 
either
by the existence of conserved global quantities with a 
back-transfer in Fourier space, or by the presence of a slow manifold 
in the dynamics.  
\end{abstract}

\maketitle

\section{Introduction}
A characteristic of turbulence is the 
dynamical involvement of fluctuations
over a broad range of length scales.
At the largest scales 
the general expectation is that 
of non-universal behavior and influence
by boundary conditions and/or driving.
At intermediate
scales, statistical behavior can be obtained,
and at the smallest scales, behavior is determined 
by some dissipation mechanism. 
Associated with these length scales, there 
are corresponding ranges of time 
scales.
The characteristic nonlinear time scale,
assuming local interactions in scale,
is obtained dimensionally as
the ratio $l/u_l$, where $l$ is a length  and $u_l$
is the corresponding typical velocity at $l$. 
Suppose we choose a particular length $L$ and its
associated time scale $L/u_L$ as the
unit of time measurement. Frequently $L$ will 
be the largest 
length scale in the
system, or perhaps the size of the energy-containing eddies. 
In these units
the range of locally computed 
time scales $l/u_l$ will typically 
extend from unity (or times of order one) to 
smaller time scales at the smaller length scales. 
However, there are some situations in which 
there is an emergence of much longer time scales,
which cannot be associated with 
this kind of 
an estimate 
that is local in scale. 
Some of these issues have previously been identified  
in numerical simulations \cite{DmitrukMatthaeus07},  for example in the case of 
a driven-dissipative three-dimensional (3D) magnetohydrodynamic (MHD) 
system in presence of a strong background magnetic field, and in 
two-dimensional (2D) hydrodynamic (HD) and MHD driven-dissipative systems. 
In those three cases, the presence of long time fluctuations is 
connected 
with the 
appearance of a $1/f$ noise type frequency 
power spectrum at very low frequencies $<< u_L/L$.
In many situations
a $1/f$ noise spectrum arises from 
the existence of multiple correlation times in the system 
(see \cite{vanderZiel50}). Such $1/f$ spectra associated
with long time scale fluctuations have been observed in a wide range of situations
\cite{Machlup,MontrollShlesinger82} but the presence of $1/f$ noise in  turbulent flows
was reported only recently \cite{DmitrukMatthaeus07}.
Dynamo simulations of the generation of magnetic fields also exhibit
$1/f$ spectra \cite{PontyEA04}.
Low frequency 
dynamics have also been reported in laboratory experiments of HD and 
MHD flows \cite{delaTorre07,Monchaux09}.
Closely related are observations in the solar wind of 
$1/f$ 
behavior in the spacecraft frame magnetic and density 
spectra at low frequencies
\cite{MatthaeusGoldstein86,MatthaeusEA07}.
However in the solar wind case the $1/f$ signal appears to 
trace back to the corona or even to the photosphere \cite{BemporadEA08}.
It is therefore a consequence of long time scale variability 
of the solar wind sources, rather than a property
that emerges due to solar wind dynamics itself.

It is conjectured in \cite{DmitrukMatthaeus07} that 
the long time fluctuations in the turbulent MHD flows 
arise from the non-local couplings between the longest length scale 
in the system (the $k=1$ mode in Fourier space) with the smaller scales, 
from the inertial to the dissipation range ($k \gg 1$). Such non-local 
interactions are known to be stronger for MHD than for HD flows 
\cite{Alexakis05,Mininni11}. It remains however an open issue as  to whether 
the existence of dissipation, the effects of the driving, or the 
particular choice of boundaries, are 
influential in determining the character or presence of the $1/f$ noise
that is generated.
The present study  attempts to show that the presence of 
a $1/f$ 
noise spectral regime is 
specifically with the structure of the equations 
and the non-linear couplings.
We accomplish this by employing 
a series of numerical simulations of ideal flows in a variety of situations. 
We emphasize that these simulations are neither
dissipative, nor driven.
Furthermore, the $1/f$ regime
will be characterized by showing its existence (or absence), and its
range of frequencies, 
for several different kinds of ideal flows
and geometries.
We will argue that 
a key feature in each system
is the existence of either 
ideal global invariants, or quasi-invariants (to be defined below), 
which provide constraints on 
behavior of a small number of important degrees of 
freedom in the system. 

We will further argue below that 
a related issue 
concerning ideal flows is the
phenomenon of delayed ergodicity \cite{ServidioEA08}, also identified 
as broken ergodicity \cite{Shebalin89,Shebalin10}. In this phenomenon, 
the ideal flow is described as a statistical mechanical system composed 
of Fourier modes (degrees of freedom).
Some of these Fourier
modes (those corresponding to the largest length scale in the system) 
appear to spend long periods of time in restricted regions of phase space
(i.e., in some interpretations, 
thus breaking the ergodicity assumption in statistical
mechanical systems). 
As it will be clear from the present study, this effect
is essentially the same as that which produces the $1/f$ spectra.

Additionally, we make a further 
connection with another phenomenon observed on long
times scales -- the spontaneous flipping or reversal of some 
large scale fields. A prominent example is 
the reversal of magnetic field or its magnetic dipole in 
MHD systems, 
an effect well known in the context  of geomagnetic studies 
\cite{GlatzmaierRoberts95,Benzi05,Sorriso-ValvoEA07}.

The organization of the paper is as follows. In section II the model
equations, 
the standard ideal equations of
HD and MHD flows, are introduced, 
and the numerical method to solve 
them is described. In section III the results are presented, divided 
into subsections for different kinds of systems. In section IV a 
discussion is developed, followed by the conclusions and summary 
in section V.

\section{Model equations}

In this paper we consider several systems of equations, including 
HD flows with and without rotation, and MHD flows with and without 
externally imposed magnetic fields. We also consider an approximation 
of the MHD equations in the limit of strong imposed magnetic fields, 
the so-called reduced MHD (RMHD) equations. All systems are 
three-dimensional, 
and the flows are incompressible
and ideal (zero dissipation coefficients).

In the most general case, the incompressible 
HD and MHD equations can be written in dimensionless 
form as
\begin{equation}
   \frac {\partial {\bf u}}{\partial t} +
   {\bf \omega} \times {\bf u} +
   2 {\bf \Omega} \times {\bf u} = -\frac{1}{\rho}\nabla {\cal P} +
   {\bf j} \times {\bf B} ,
\label{eq:Euler}
\end{equation}
\begin{equation}
  \frac{\partial {\bf B}}{\partial t} = \nabla \times ({\bf u} \times 
  {\bf B}) ,
\label{eq:induction}
\end{equation}
where ${\bf B}$ is the magnetic field, ${\bf u}$ the velocity field, 
${\bf j} = \nabla \times {\bf B}$
the current density, ${\bf \omega} = \nabla \times {\bf u}$ 
the vorticity, and ${\cal P}$ the pressure.
A term considering rigid rotation with angular velocity
$\Omega$ can be included in the velocity equation. This term corresponds
to the Coriolis force, with the centrifugal force absorbed into
the total pressure term. The pressure 
can be obtained taking the divergence of the velocity equation, 
using the incompressibility condition $\nabla \cdot {\bf u} = 0$,
and solving the resulting Poisson equation.
The solenoidal ($\nabla \cdot {\bf B} = 0$)
magnetic field includes a uniform part ${\bf B_0}$ and a fluctuating
part ${\bf b}$, so that ${\bf B} = {\bf b} + {\bf B_0}$.

With ${\bf B_0}=0$, these equations reduce to the MHD equations without 
an external field. When ${\bf B}\equiv 0$, the equations are the ideal 
incompressible  three-dimensional hydrodynamic equations, that is, 
the Euler equations, which we also consider as a case study. Finally, 
when ${\bf \Omega} \neq 0$ any of these systems is written in a rotating 
frame, while ${\bf \Omega} = 0$ corresponds to the non-rotating case. 

The RMHD equations
can be derived from Eqs. (\ref{eq:Euler}) and 
(\ref{eq:induction}) for ${\bf \Omega} = 0$ and in the limit of strong 
${\bf B_0}= B_0 \hat {\bf z}$, assuming low frequencies 
and weak 
spatial gradients 
along the direction of the background magnetic field 
\cite{Strauss76,Montgomery82}.
The equations
involve potentials
$a(x,y,z,t)$ and $\psi(x,y,z,t)$
such that in rectangular $(x,y,z)$ coordinates
one has 
${\bf b} = \nabla_\perp \times \hat {\bf z} a$ and 
${\bf u} = \nabla_\perp \times \hat {\bf z} \psi$, 
where $\nabla_\perp = (\partial_x, \partial_y,0)$.
To get slow dynamics as ${\bf B}_0 \sim 1/\epsilon \to \infty$,
for small $\epsilon$,
one is forced to an ordering such that $\partial_z = O(\epsilon)$.
In this sense the RMHD equations are ``quasi-two dimensional'' in 
$x$ and $y$. The dynamical equations become simply those of the 
potentials,
\begin{eqnarray}
\frac {\partial \omega}{\partial t} + {\bf u} \cdot \nabla_\perp \omega
= & {\bf b} \cdot \nabla_\perp  j + B_0 \frac {\partial j}{\partial z} + \mu \nabla^2 \omega
\nonumber
\\
 \frac{ \partial a}{\partial t} + {\bf u} \cdot \nabla_\perp a = &
B_0 \frac {\partial \psi}{\partial z} + \mu \nabla^2 a
\label{eq:rmhd}
\end{eqnarray}
where 
the electric current density is $j=-\nabla^2 a$ and 
the vorticity is $\omega = -\nabla^2 \psi$.

In the subsequent section we will describe numerical results obtained with two different
type of codes, which we now describe briefly.
The parameters of 
all runs discussed below are given in Table I for reference.

For one set of numerical simulations, we assume periodic boundary conditions
in a cubic box, of side $2\pi L_0$ in each cartesian direction, with $L_0$
the arbitrary unit of length. 
Fields can then be decomposed in Fourier modes
\begin{equation}
{\bf u} = \sum_{\bf k} {\bf u}_{\bf k}
               {\rm e}^{i ~{\bf k} \cdot {\bf r}} ~,~
{\bf b} = \sum_{\bf k} {\bf b}_{\bf k}
               {\rm e}^{i ~{\bf k} \cdot {\bf r}} ~,
\end{equation}
where  ${\bf u}_{\bf k}$, ${\bf b}_{\bf k}$ are the Fourier coefficients of the expansion,
and ${\bf k}$ is the wavevector, having 
integer components in 
the dimensionless case.

We employ 
a pseudospectral code to accurately numerically solve these
equations. With the pseudospectral method (using full de-aliasing with
the 2/3 rule) any quadratic invariant (like the total energy) 
is exactly maintained, except for machine round-off errors and 
time integration discretization errors. 
Time integration here is done with a second-order Runge-Kutta method,
with a very small time step $dt$, 
to control the discretization error over the long simulations
we carry out. 
Typically we use $dt \sim 5 \times 10^{-4} $
which is much less than the global large scale turnover time
$L_0/u$, where $u$ is the root mean square (rms) velocity. 
As an example, in an integration of 
1000 unit times duration, for initial primitive fluctuation 
fields $\bf u$ and $\bf b$ with rms values of 1, these  
values remain equal to unity with an error less than $5 \times 10^{-6}$ at 
the end of the integration. In dissipative MHD
the energy $E$ balance equation satisfies

\begin{equation}
\frac{dE}{dt} = -2 \nu (< j^2 > + <\omega ^2>)
\end{equation}

The decay in energy due to time discretization errors can be
interpreted as numerical dissipation and for the example 
mentioned, using a time averaged value for 
$<j^2> + <w^2> \sim 10$,  the numerical viscosity 
is estimated as $\nu \sim 10^{-10}$. 

In a second set of numerical simulations, we consider spherical geometry: the 
HD or MHD equations are solved inside a sphere of unit dimensionless 
radius, with 
vanishing velocity and magnetic field at the sphere boundary.
For this geometry, a fully spectral Galerkin code is
used, based on a Chandrasekar-Kendall (C-K) decomposition
of the fields. The C-K functions \cite{Chandrasekhar57,Montgomery78}
are 
\begin{equation}
{\bf J}_i = \lambda \nabla \times {\bf r} \psi_i + \nabla \times \left(
    \nabla \times {\bf r} \psi_i \right) ,
\label{eq:ji}
\end{equation}
where we work with a set of spherical orthonormal unit vectors
$(\hat{r},\hat{\theta},\hat{\phi})$, and the scalar function $\psi_i$ is a
solution of the Helmholtz equation, $(\nabla^2 + \lambda^2) \psi_i = 0$.
The explicit form of $\psi_i$ is
\begin{equation}
\psi_i (r, \theta, \phi) = C_{ql} \, j_l(|\lambda_{ql}| r) Y_{lm}
    (\theta,\phi),
\label{eq:psi}
\end{equation}
where $j_l(|\lambda_{ql}| r)$ is the order-$l$ spherical Bessel function of the first
kind, $\{ \lambda_{ql}\}$ are the roots of $j_l$ indexed by $q$
(so that the function vanishes at $r=1$), 
and $Y_{lm}(\theta,\phi)$ is a spherical
harmonic in the polar angle $\theta$ and the azimuthal angle
$\phi$.
The sub-index $i$ is a shorthand notation for the three indices $(q,l,m)$;
$q=1,2,3,\dots$ corresponds to the positive values of $\lambda$, and
$q=-1,-2,-3,\dots$ indexes the negative values; finally $l=1,2,3,\dots$, 
and $-l \le m \le l$. The C-K functions satisfy
\begin{equation}
\nabla \times {\bf J}_i = \lambda_i {\bf J}_i  \ .
\end{equation}
With the proper normalization constants, they are an orthonormal set that
has been shown to be complete \cite{Cantarella00}. The values of 
$|\lambda_i|$ play a role similar to the wavenumber $k$ in the Fourier 
expansion. Note that boundary conditions, as well as the Galerkin method to 
solve the equations inside the sphere using this base, were chosen to 
ensure conservation of all quadratic invariants of the systems, crucial 
for our present study of ideal flows for long times. More details about the 
technique to numerically 
solve the HD and MHD equations in this spherical geometry can be found 
in \cite{Mininni06,Mininni07}.

\section{Results}

\subsection{Three-dimensional MHD in a box and in the sphere}

For three-dimensional incompressible ideal MHD
with no mean magnetic field, 
there are three quadratic
invariants: the total (kinetic plus magnetic) 
energy per unit mass
\begin{equation}
E=\frac{1}{2} \langle |{\bf u}|^2 + |{\bf b}|^2 \rangle \ = \ E_u+E_b ,
\end{equation}
(with $\langle \dots \rangle$ denoting a spatial average),
the cross helicity
\begin{equation}
H_c= \langle {\bf u} \cdot {\bf b} \rangle \,  ,
\end{equation}
and the magnetic helicity
\begin{equation}
H_m= \langle {\bf a} \cdot {\bf b} \rangle \,  
\end{equation},
where ${\bf a}$ is the 
vector potential such that $\nabla \times {\bf a} = {\bf b}$.
The robustness of the Gibbs equilibrium ensemble predictions for 
this system is well established \cite{Lee52,Kraichnan65,Shebalin83,FrischEA75,StriblingMatthaeus90}.
The expectation value of 
spectra are readily obtained in the Gibbs ensemble
using Lagrange multipliers associated with each conserved
quantity. These 
expectations 
are well verified with numerical simulations \cite{FrischEA75,StriblingMatthaeus90}.
Usually the Gibbs ensemble has been viewed as a predictor
of the direction of spectral transfer,
although more recently
it has also been found that the 
ideal Gibbs-Galerkin system 
shares additional characteristics with the dissipative turbulent system
at short and intermediate time scales and length scales.
One of the characteristics found is that even in the ideal 
truncated case, cascades develop during transients 
(\cite{brachet, KrstulovicEA09, WanEA09, giorgio}).
For long times the system goes to
solutions with zero flux but
any perturbation of the system away from the zero
flux solutions (e.g., by thermal fluctuations) is corrected by 
transient non-zero fluxes associated with the non-linear interactions
(\cite{Mininni11DMP}).

For wavenumber $k >> 1$ 
(the wavenumber $k$ in the following discussion 
should be considered equivalent to $|\lambda_i|$ for the spherical case),  
the Gibbs ensemble predicts in three-dimensions an omnidirectional
spectrum going 
like $k^2$. Attaining this equilibrium prediction
is frequently called ``thermalization'' of the large wavenumber modes, 
as this 
corresponds to equipartition of energy among all individual modes, i.e.,
a flat modal spectrum. 
At the fundamental modes $k=1$ (the largest possible wavelength modes 
in a finite size system), condensation is predicted, according to the 
established value of $H_m$. When $H_m \neq 0$, condensation at the
$k=1$ mode occurs, and this has been the base for prediction \cite{FrischEA75}
of an 
inverse cascade of magnetic helicity in dissipative MHD (i.e., when
dissipation and forcing is added to the ideal MHD equations).

Furthermore, when both $H_m \neq 0$ and 
$H_c \neq 0$, the condensation of magnetic helicity induces a partial condensation 
of the cross helicity \cite{StriblingMatthaeus90}. For such cases the 
largest scales are expected to contain
signatures both of magnetic helicity (helical ${\bf b}$)
and of Alfv\'enic
correlation (${\bf u} \propto {\bf b}$).

We note here that 
the ideal model can be viewed as a dynamical
model of the nonlinearities that drive turbulence. It is a
simplified model because it becomes Gaussian, and also it 
lacks a preferred
average direction of spectral transfer. But it is worth remembering that
in ``real turbulence'', with dissipation, there are large numbers of 
couplings that take
energy to higher $k$, and also large numbers of couplings that take 
energy to lower $k$. These two types of couplings are almost in 
balance. However, transfer to higher $k$ dominates slightly -- this is
just what the direct cascade is (an inverse cascade would be the
case where the transfer to lower $k$ dominates slightly).
This has been observed in numerical simulations, as well as in 
laboratory experiments.
In this regard, the ideal model is not ``real turbulence'' but it
shares some of its properties (\cite{brachet}).

An assumption for the equilibrium ensemble predictions is the 
property of ergodicity. The required property
is that, in a long period 
of time, the system, defined by the set of real and imaginary parts of its
Fourier coefficients (i.e., the dynamical degrees of freedom of 
the ideal MHD system) will visit all accessible regions
in complex phase space. A point in phase space is ``accessible'' 
if it is permitted
by the values of the invariant quantities.
This is well verified for $k > 1$ modes, 
which evolve on relatively fast time scales.
However, 
for the longest wavelength $k=1$ modes 
there is an apparent breaking of ergodicity that has been 
identified in simulations \cite{Shebalin89,Shebalin10}.
There is however evidence that ergodicity
is restored at 
very long times \cite{ServidioEA08}. 
As a result, the $k=1$ 
modes seem to wander for very long periods of time in restricted regions 
of phase space, until at some point, a possibly sudden
``hopping'' 
\cite{ServidioEA08} occurs  
and a new period of wandering occurs 
in another restricted region of phase space. As has been recently 
also pointed out \cite{Shebalin10}, the time duration
of these wandering periods is related to the
dimension of the system (i.e., the finite number of modes assumed for
the numerical simulation). 
This is to be expected since the 
delayed ergodicity is related to condensation, and 
for any system with the same values of the ideal invariants,
the 
condensation
is more complete for systems with larger numbers of degrees of freedom 
 \cite{StriblingMatthaeus90}.

We will focus here then on the time
behavior of a single $k=1$ mode, and the point we make is that
there is a connection between 
this 
aperiodic cycle of wandering and hopping behavior, and the $1/f$
power frequency spectrum already observed in dissipative MHD 
\cite{DmitrukMatthaeus07}. 
The time behavior of the $k=1$ mode
can be interpreted as the time behavior of the large scale magnetic 
field, i.e., it corresponds to the time behavior of the 
fluctuating magnetic field after a filtering of the smaller scales 
(with $k > 1$) is performed. This connection between $k=1$ and the 
apparent or observable large scale magnetic field becomes sharper 
as the condensation becomes more complete. 
In dissipative MHD, this filtering process 
occurs naturally through dissipation which damps the small scales. 
In ideal MHD, however, the thermalization of the small scales, 
with a spectrum going like $k^2$,
 tends to obscure the time behavior of 
the $k=1$ mode if a power frequency analysis is applied to the 
full fluctuating magnetic field. As a result, we shall look at the time 
behavior of a single $k=1$ mode in this case, in order to properly see the
long time fluctuations that develop. In what follows, we thus present
a series of results for the time behavior and the frequency
spectrum of the $k=1$ mode for different numerical simulations
of ideal MHD, with first a fixed system size (fixed $N$, with $N^3$  = total number of
modes) and varying the magnetic helicity, and secondly by varying 
the number of modes $N^3$ but with approximately fixed magnetic helicity. 
Our working hypothesis is that these scalings can be understood 
in terms of scalings within the Gibbs ensemble \cite{StriblingMatthaeus90}:
For fixed $H_m/E$, increasing $N$ intensifies the condensation until all $H_m$
asymptotically resides in $k=1$. For fixed $N$, increasing $H_m/E$ increases 
condensation until as this ratio approaches its maximum value, all excitations
condense to $k=1$.

\subsubsection{Periodic box} \label{ss:PB}

\begin{figure}
\epsfysize=16cm
\centerline{\epsffile{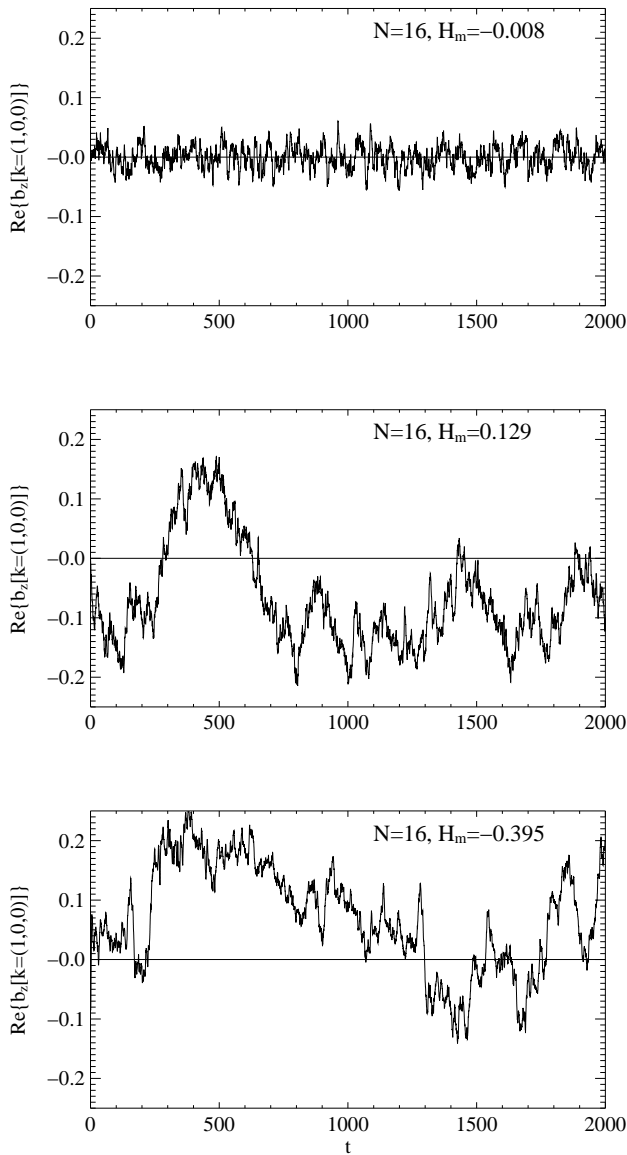}}
\caption{
Time series of a component of the magnetic field
for the ${\bf k}=(1,0,0)$ mode in 
a MHD run  with $16^3$ modes and with different values of the
magnetic helicity: $H_m=-0.008$ (top), $0.129$ (middle), and $-0.395$ (bottom).
}
\label{fig:timeseriesHm}
\end{figure}

We begin by considering the behavior in time of 
a large scale mode in a sequence of three  incompressible ideal 3D MHD 
simulations with 
increasing $|H_m|$. These are further described in Table I,
and have 
$H_m=-0.008$, $0.129$, and $-0.395$,
a fixed simulation size of $N^3=16^3$, and 
fixed total energy equal to $1$.
The fluctuations are initially equipartitioned 
$E_u=E_b$,
and concentrated 
in a range of wavenumbers $1 < k < 4$.
In particular
Figure (\ref{fig:timeseriesHm})  illustrates  
the temporal behavior of the $k=1$ modes, 
choosing the real part of $b_z ({\bf k}=1,0,0)$ as
an indicator of the 
behavior of the large-scale modes.
Indeed, the same behavior is obtained for other components, 
$b_x$, $b_y$, or for other directions in ${\bf k}$, except of course 
for directions for which ${\bf k}$ is parallel to the field component, 
which are identically zero due to the $\nabla \cdot {\bf b}=0$ condition. 
The different panels in Figure (\ref{fig:timeseriesHm}) 
correspond to increasing values
of $|H_m|$. 
It is apparent that as the magnitude of 
magnetic helicity is increased, 
long period fluctuations (as long as 
$1000$ unit times) begin to appear.
Such fluctuations are not observed in the time series of modes with 
larger wavenumber (i.e., smaller spatial scales), and 
are characteristic of the largest scale Fourier modes in the system. 
Note that 
the value of the magnetic field fluctuation amplitude
increases with $H_m$, consistent with the condensation phenomenon.

\begin{figure}
\epsfysize=17cm
\centerline{\epsffile{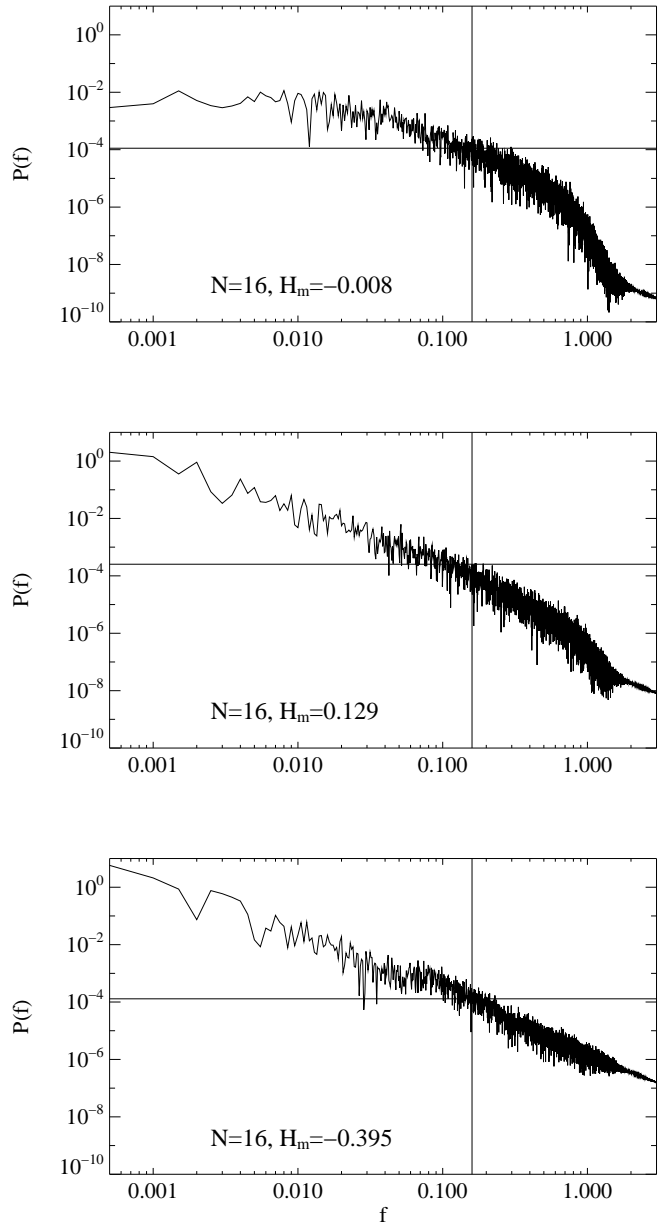}}
\caption{
Power frequency spectra of the time series 
in Figure (\ref{fig:timeseriesHm}), with different values of the
magnetic helicity $H_m=-0.008, 0.129$, and $-0.395$.
The reference vertical line corresponds to a frequency value $f_L=1/(2\pi)$
(see text) and the horizontal line to the value of $P(f_L)$.
}
\label{fig:freqspectrHm}
\end{figure}

Qualitatively, it would be reasonable to
say that the low helicity case in Figure (\ref{fig:timeseriesHm}) (top)
is more ``stationary'', as the fluctuations 
in that case approach a zero mean within say 100 time units
or less.
On the other hand the two higher helicity cases exhibit coherent 
fluctuations at a scale of hundreds, or even thousands,
of time units. The suggestion is that low frequency
oscillations are becoming more dominant with higher helicity.
Indeed,
there is clear evidence of this
in the frequency 
spectra, shown in Figure (\ref{fig:freqspectrHm})  
for each of the three time series given in Figure (\ref{fig:timeseriesHm}).
The increasing power in the low frequency part,
and the emergence of a $\sim 1/f$ power law
at the very low frequencies $<<1$ , 
are associated with increasing $|H_m|$.
We note here that often in the 
literature, $1/f$ is loosely used to refer to any spectrum 
of the form $f^{-\alpha}$, with $0<\alpha<2$ 
(i.e. omitting both white noise and Brownian motion). 

The important point that we want to stress here is not
the exact power law index that fits the spectra (which in
fact is dependent on the time duration of the simulations,
because when longer time fluctuations appear a more extended
run would be needed), but rather the fact that there is 
no obvious reason for which fluctuations with time scales
orders of magnitude longer than the unit time should appear 
here. The longest time scale based on local time arguments
is $T_L=L/u_L$, which for the largest length scale of $L=2\pi$ (box size)
corresponding to $k=1$ and for $u_L=1$, is $T_L=2\pi$. 
Frequencies below $f_L = 1/T_L = 1/(2\pi)$ should normally
have a flat power spectrum (white noise), indicating the
existence of a defined correlation time. However as the
Figure (\ref{fig:freqspectrHm}) shows, 
for the cases where long time fluctuations appear,
the spectrum to the left of the $f_L= 1/(2\pi)$ is far
from being flat. This is indicated in the panels of the figure
with a vertical line at the frequency $1/(2\pi)$ and a horizontal line
at the corresponding value of the power spectrum for that frequency. 
We identify the difference between the observed power spectrum and
a flat power spectrum as the range of $1/f$ spectra in each case
(again, using a loose definition for $1/f$).

This range of the $1/f$ spectra is short for the lower
values of $H_m$, with corresponding time periods ($T \sim 1/f$) of the 
order of $10$ unit times, and 
increasing to $T \approx 100-500$ for $H_m=0.129$, and as long
as $T \approx 500-1000$ for $H_m=0.395$.

It is interesting to observe the autocorrelation function for each of the
time series in the cases shown, as a complementary way to see the
appearance of long time fluctuations. 
Here the autocorrelation function is defined as 
\begin{equation}
C(t) = < b(t_0) b (t_0+ t) > 
\end{equation}
where $b$ represents a cartesian component of a magnetic field mode 
(for instance
the $z$ component of the real part of the magnetic field Fourier mode 
for $k=(1,0,0)$),
with subtracted mean and normalized so that $<b^2>=1$,
$t_0$ is an arbitrary origin in time, $t$ is the time lag
and $< ... >$ denotes a time average.

\begin{figure}
\epsfysize=17cm
\centerline{\epsffile{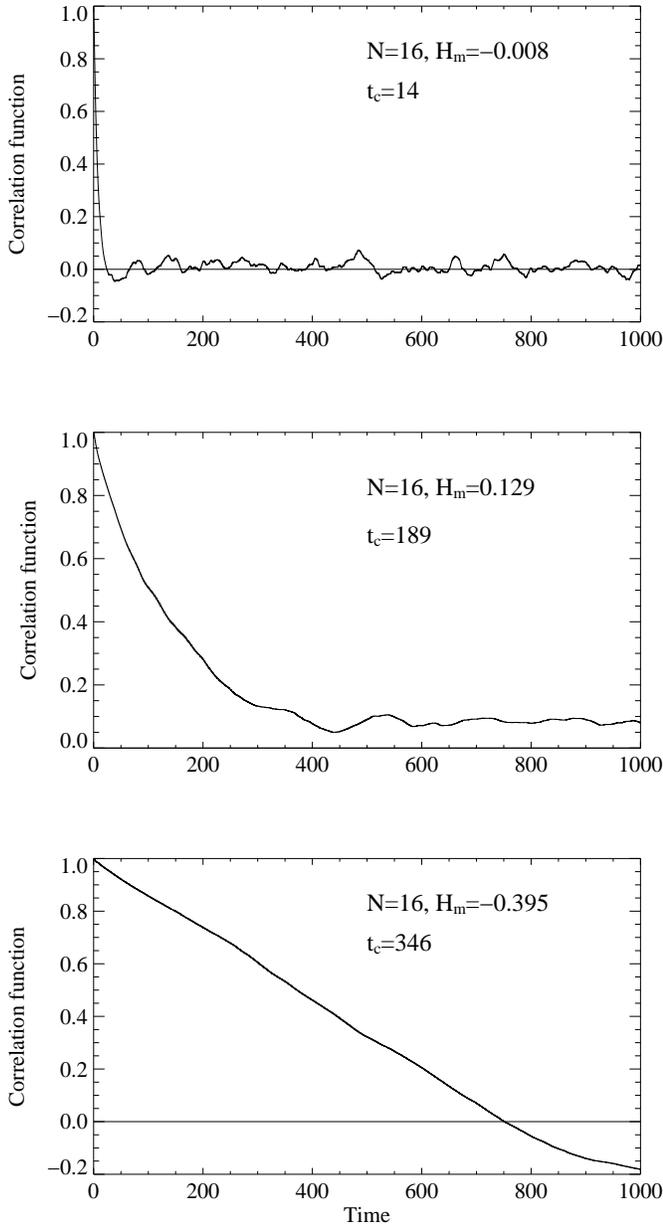}}
\caption{
Self-correlation function for the time series of magnetic field in 
Figure (\ref{fig:timeseriesHm}), with different values of the
magnetic helicity $H_m=-0.008, 0.129$, and $-0.395$.
A correlation time $t_c$ (see text) is indicated in each case.
}
\label{fig:correlfunctHm}
\end{figure}

This is shown in 
the panels of Figure (\ref{fig:correlfunctHm}). 
The case with smallest $H_m$ presents a compact
correlation function, localized within about $10$ unit times, whereas the 
cases with larger $H_m$ show a much broader correlation function.
A correlation time can be obtained as
\begin{equation}
t_c = \int_0^{T_f} C(t) dt
\end{equation}
where $T_f$ is the final time of the run, assuming it is a large time.

This quantity is well defined if the correlation function is confined
within a finite range of time lags,
decaying faster for long times. However, if the correlation function
does not decay faster, and in depends on the time duration of the
series, then the correlation time is not well defined. In particular, for an
exact $1/f$ power spectrum there is no single correlation time
that can be defined. 
The values indicated in the plot for the three $H_m$ cases
of $t_c=14, 189, 346$ show again that longer time fluctuations (and
correlations) appear as $H_m$ is increased.

\begin{figure}
\epsfysize=17cm
\centerline{\epsffile{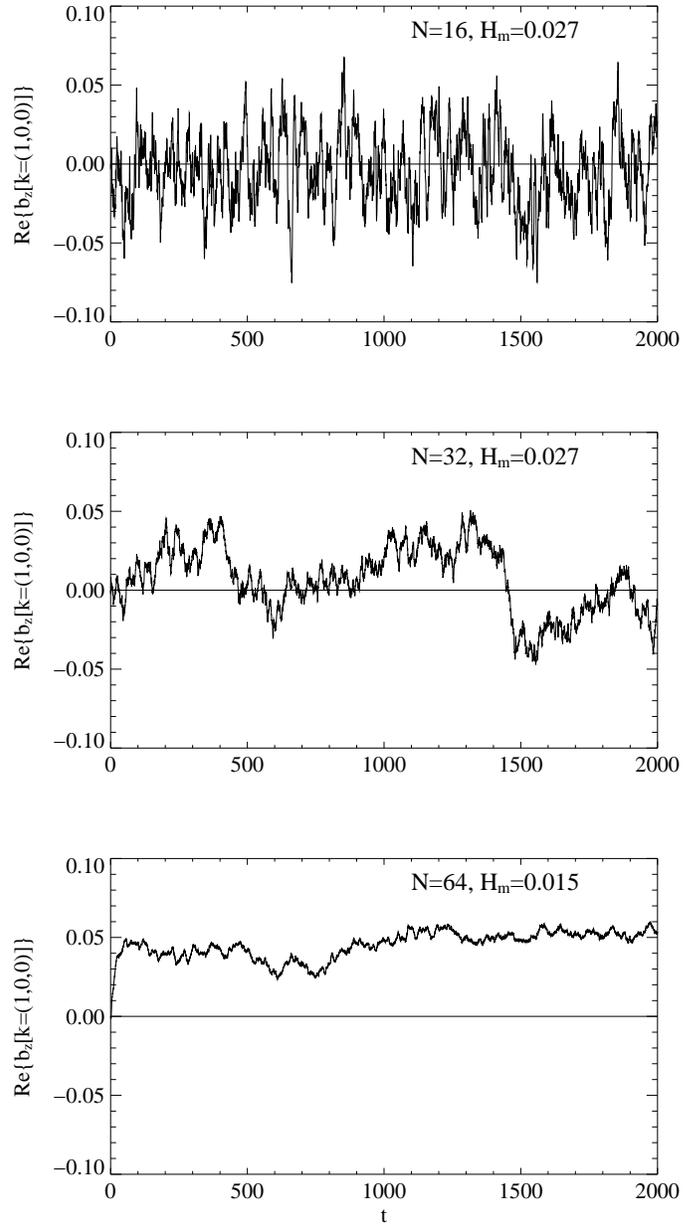}}
\caption{
Time series of a component of the magnetic field
for $k=(1,0,0)$ mode in 
the MHD runs with different resolution $N^3=16^3, 32^3$, and $64^3$
and magnetic helicity $H_m=0.027, 0.027$, and $0.015$.
}
\label{fig:timeseriesN}
\end{figure}

\begin{figure}
\epsfysize=17cm
\centerline{\epsffile{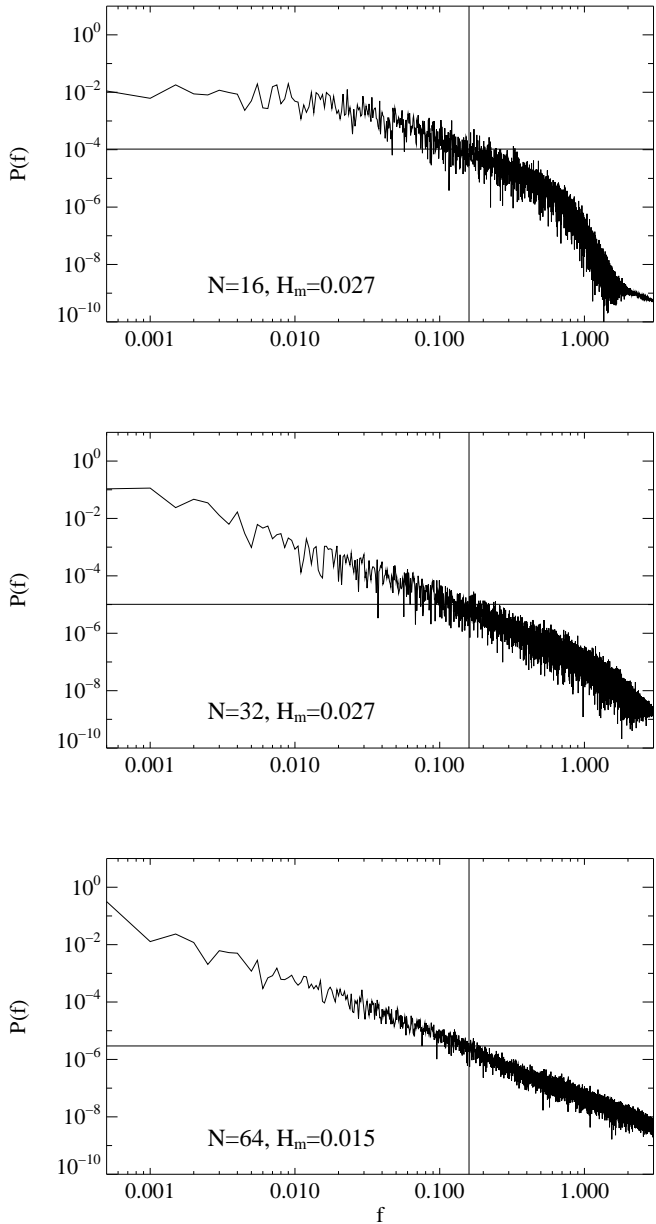}}
\caption{
Power frequency spectra of the time series 
in Figure (\ref{fig:timeseriesN}), 
with different resolution $N^3=16^3, 32^3$, and $64^3$ and
magnetic helicity $H_m=0.027, 0.027$, and $0.015$.
The reference vertical line corresponds to a frequency value $f_L=1/(2\pi)$
(see text) and the horizontal line to the value of $P(f_L)$.
}
\label{fig:freqspectrN}
\end{figure}

Recalling that condensation in the ideal Gibbs ensemble 
intensifies as the number of degrees of freedom ($\sim N^3$) increases,
we now explore whether the emergence of long time scale coherent
fluctuations and the associated $1/f$ spectra 
behave the same way. 
The next series of plots in Figure (\ref{fig:timeseriesN}) 
show the time behavior 
of a magnetic field component of a single mode ${\bf k}=(1,0,0)$ 
varying the size of the simulation of 3D MHD, for $N^3=16^3, 32^3$, and 
$64^3$, and for low values of $H_m=0.027$, $0.027$, and $0.015$, 
respectively.
for each $N$. These results show that long time fluctuations
appear even for low values of $H_m$ when the size of the system
is increased. We note that for the $N=64$ case a longer time range
is needed to observe fluctuations to average to zero. This is indeed
observed in this run when it is extended until $t > 10000$ but not
shown here for consistency with the shorter time range selected 
for the rest of the cases in this Figure. 
Figure (\ref{fig:freqspectrN}) shows the corresponding power frequency 
spectra $P(f)$ for each size $N$. The range of low frequency $1/f$ 
noise increases from $T \sim 10$ for $N=16$ to $T \sim 2000$ for
$N=64$ (compare Figures (\ref{fig:freqspectrHm}) and (\ref{fig:freqspectrN})).

\begin{figure}
\epsfysize=17cm
\centerline{\epsffile{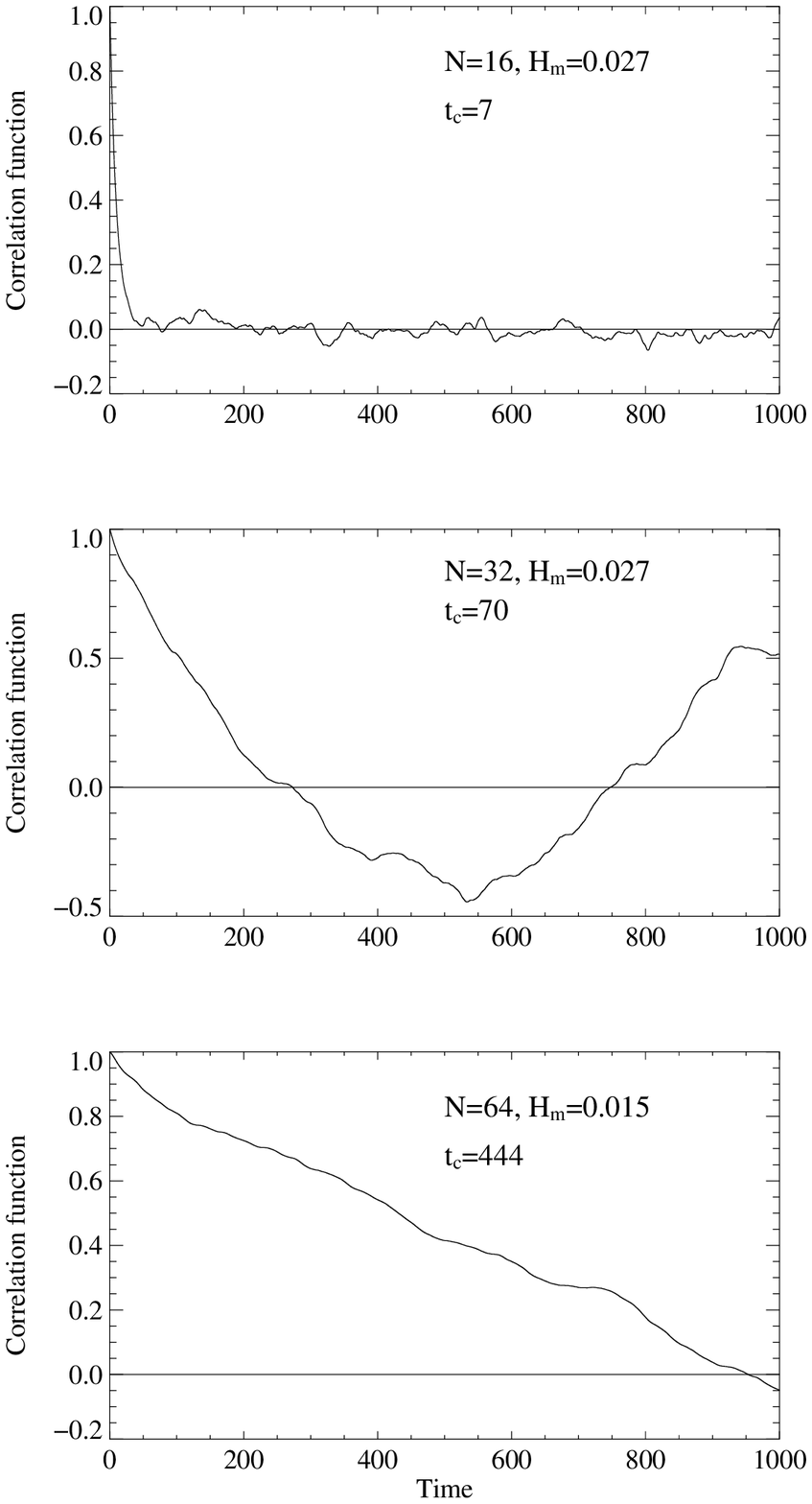}}
\caption{
Self-correlation function for the time series of magnetic field in 
Figure (\ref{fig:timeseriesN}), 
with different resolution $N^3=16^3, 32^3$, and $64^3$ and
magnetic helicity $H_m=0.027, 0.027$, and $0.015$.
A correlation time $t_c$ (see text) is indicated in each case.
}
\label{fig:correlfunctN}
\end{figure}

Furthermore, Figure (\ref{fig:correlfunctN}) 
shows the autocorrelation function 
and the obtained correlation time $t_c$ for each
time series for different values of $N$. Also evident in this
figure is the appearance of long time fluctuations as $N$ is increased
through the broadening of the autocorrelation function.

\begin{figure}
\epsfysize=9cm
\centerline{\epsffile{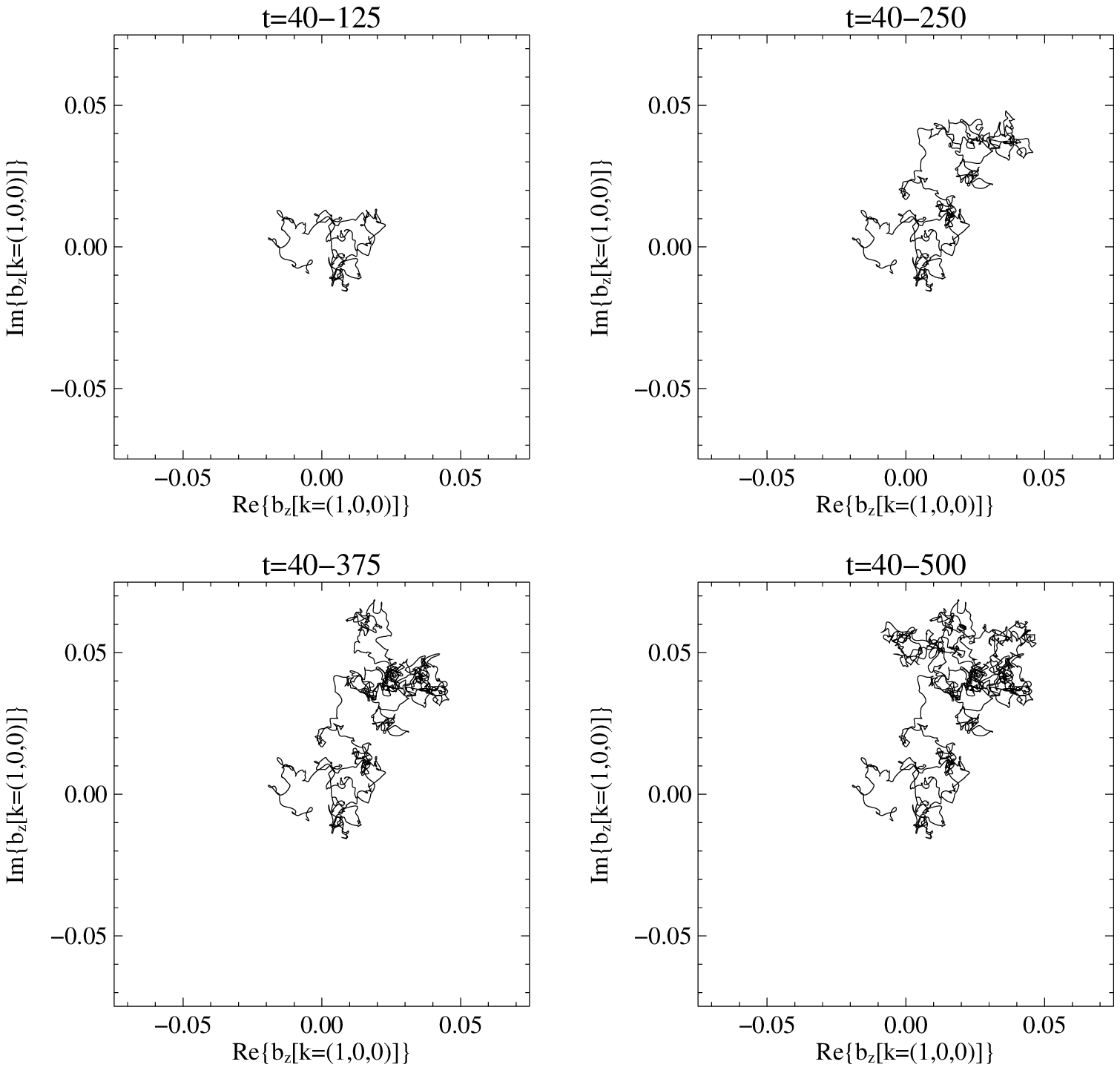}}
\caption{
Complex phase space trajectory for the ${\bf k}=(1,0,0)$ mode in the MHD 
$32^3$ run, with $H_m=0.027$,
for different intervals of time.
}
\label{fig:pspace100}
\end{figure}

\begin{figure}
\epsfysize=9cm
\centerline{\epsffile{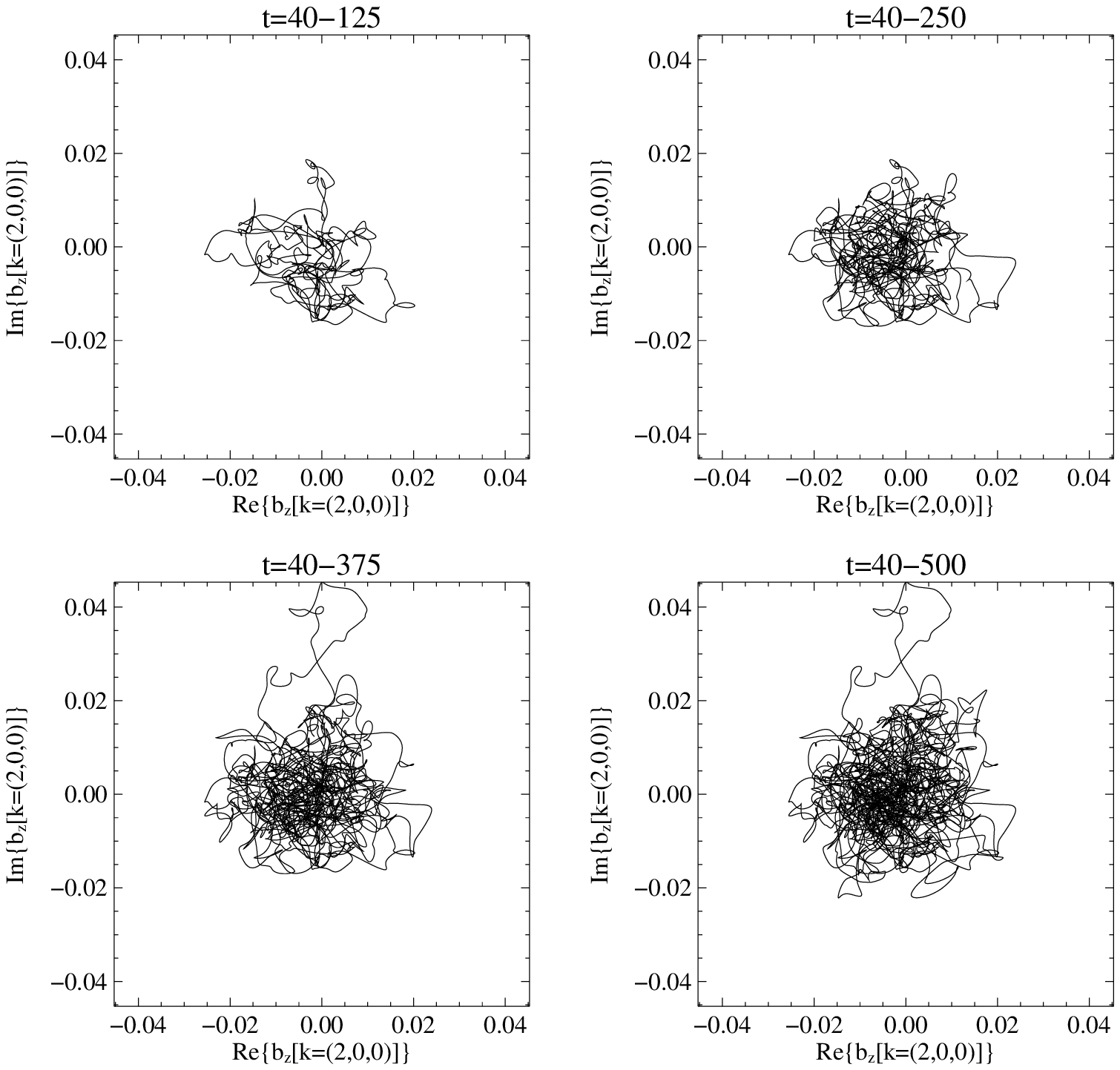}}
\caption{
Complex phase space trajectory for the ${\bf k}=(2,0,0)$ mode in the MHD 
$32^3$ run, with magnetic helicity $H_m=0.027$. 
Notice the noticeably larger wandering in phase space when compared to 
Figure (\ref{fig:pspace100}), indicative of clearer ergodicity.
}
\label{fig:pspace200}
\end{figure}

Another view of the long time fluctuations can be obtained from the 
time evolution of individual Fourier modes in a phase space plot 
in the complex plane \cite{Shebalin89,Shebalin10}. 
For instance, the $b_z(1,0,0)$ mode
for the $32^3$ run in the $H_m=0.027$ case is shown in 
Figure (\ref{fig:pspace100}).
For comparison, the behavior in the complex plane of the $b_z(2,0,0)$ mode, 
with a larger $k$, for the same run is shown in 
Figure (\ref{fig:pspace200}).
The long time fluctuations of the $b_z(1,0,0)$ mode 
correspond to long periods of time spent
in a restricted region in the complex plane, as contrasted by the 
quick filling of the complex plane allowed region for the $b_z(2,0,0)$
case. This phenomenon
has been called \cite{ServidioEA08} ``delayed ergodicity,'' 
because the ergodicity property of the $k=1$ seems to be broken only temporarily, 
as the 
mode spends long times in a region of the complex plane thus not filling 
the entire space. For longer times, the ``hopping'' of the mode between 
different regions starts filling the space. As observed here, this
corresponds to long time fluctuations of the time series, and to the
appearance of the $1/f$ power law in the power frequency spectrum.
 
As can be seen in the time series plots, 
the component of the
large scale magnetic field shown, which is a dominant contribution to the 
global magnetic field when condensation is strong, 
progresses for long periods of time without 
changing sign, and then experiences a reversal in sign, 
followed by another long 
period of time without sign change 
(see e.g., Figures (\ref{fig:timeseriesHm}) and (\ref{fig:timeseriesN})). 
We will come back to this ``reversal'' 
phenomenon in spherical runs, but it can already be seen that this is 
part of the same long time fluctuations phenomenon reported as $1/f$ noise.

\begin{figure}
\epsfysize=12cm
\centerline{\epsffile{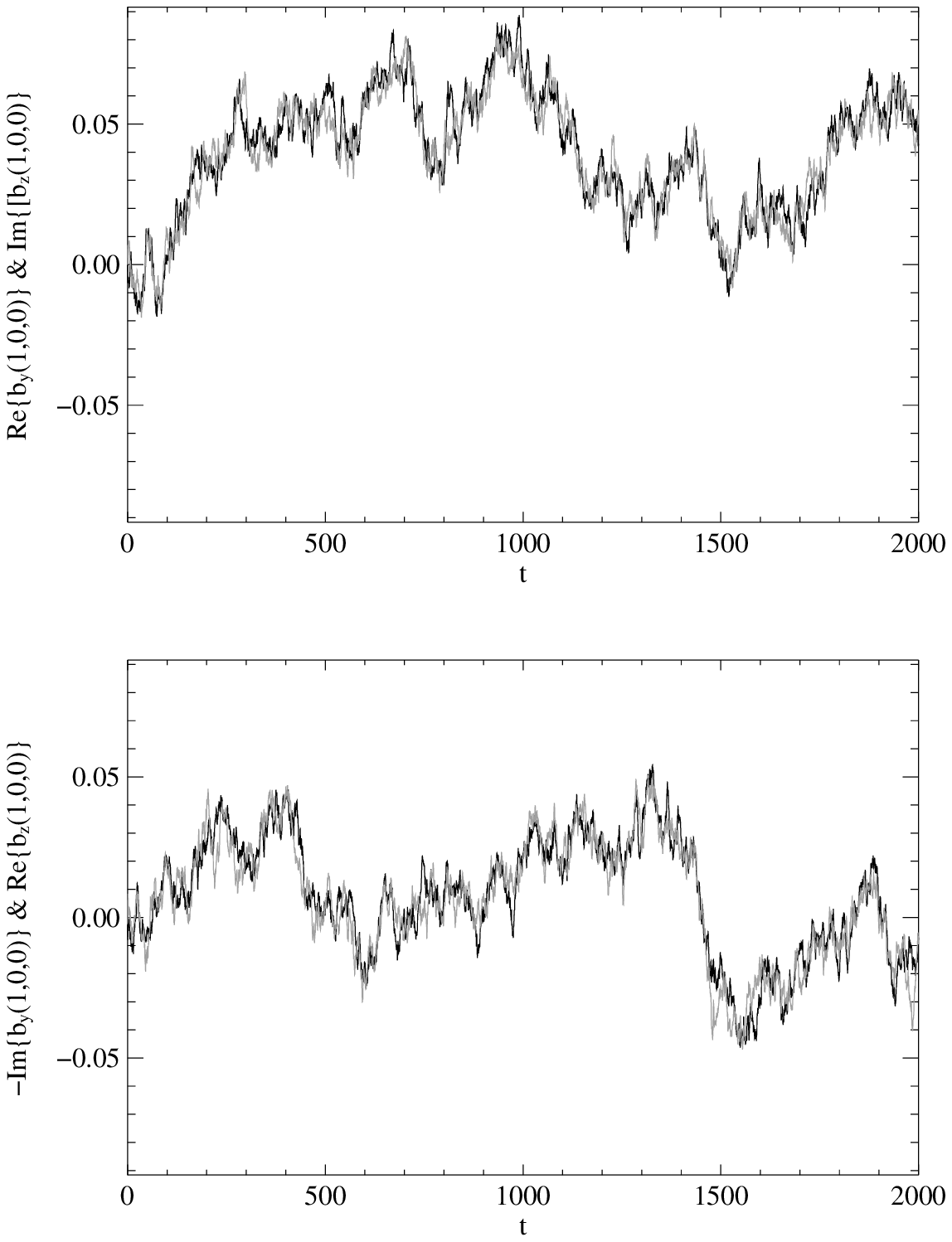}}
\caption{
Top: 
real part of the $b_y$ component of the ${\bf k}=(1,0,0)$ mode vs time (black),
and imaginary part of $b_z$ for the ${\bf k}=(1,0,0)$ mode vs
time (light gray) for the MHD $32^3$ run with magnetic
helicity $H_m=0.027$.
Bottom: 
imaginary part of $-b_y$ for the ${\bf k}=(1,0,0)$ mode vs time 
(black), and real part of $b_z$ for the ${\bf k}=(1,0,0)$ mode vs
time (light gray).
}
\label{fig:ffree}
\end{figure}

Another interesting diagnostic is shown in Figure (\ref{fig:ffree}), 
where the 
real and imaginary parts of the complex amplitudes of several field 
components of the mode ${\bf k}=(1,0,0)$ 
are shown as a function of time,
for the same run
($N^3=32^3$ and $H_m=0.027$). This is a 
case with long time fluctuations and delayed ergodicity.
Apparently, from our analysis the $k=1$ magnetic modes satisfy particular
equilibrium configurations, namely the field is quasi force-free.
A force-free magnetic field satisfies
${\bf j} \times {\bf b} = 0$, that is, the Lorentz force
term in the momentum equation is zero. If the velocity field ${\bf u}$
is zero, this means that for an ideal flow (no viscosity or diffusivity)
the magnetic field will remain force-free in
time, and all non-linear terms will be zero. Force-free states also
correspond to maximum allowable values of the magnetic helicity
(see \cite{Moffattbook}).

In terms of Fourier modes,  if a single ${\bf k}$ mode is in a 
force-free state, i.e.,
${\bf j_k} \times {\bf b_k} =0$, then, since
${\bf j_k} = i {\bf k} \times {\bf b_k}$, it must satisfy
$i {\bf k} \times {\bf b_k} = \lambda {\bf b_k}$ which implies
that $\lambda ^2 = k^2$, that is, $\lambda = \pm k$.
For the $k=1$ modes this means $\lambda = \pm 1$ and, since
these modes are equal to the Cartesian versors
(unit vectors), it also implies
some relations between the components of ${\bf b_k}$.
Taking, for example, ${\bf k}=(1,0,0)$, these relations are $Im(b_z)=\pm 
Re(b_y)$
and $Re(b_z)=\mp Im(b_y)$,
where the field components are in k-space (notice that for this mode, 
$b_x=0$
because of the divergence-free condition.)
This is immediately seen to be the condition for
a  {\it circularly polarized} fluctuation in a {\bf k}-aligned 
coordinate system --
that is, if we let ${\bf k} = (1,0,0)$ in a right-handed Cartesian 
system, then the
above conditions are equivalent to
${\bf b_k} = e^{\phi_k}(0,1,\pm i)$ for arbitrary phase $\phi$.
The plots in Figure (\ref{fig:ffree}) show
that the imaginary and real parts of these field components exhibit this 
correlation --
they are very highly
correlated in time (light gray
and black are used for each corresponding component and are almost 
indistinguishable).
In fact, values of the correlation
coefficient $ > 0.95$ are obtained in each case, corresponding to near 
maximum helicity.

Similar relations
can be found for the ${\bf k}=(0,1,0)$ and ${\bf k}=(0,0,1)$
modes, and strong correlations $> 0.95$ are also found for the
imaginary and real part of the corresponding field components
in time (not shown). This is true for all the previous runs for which
long time fluctuations of the $k=1$ modes are observed.
However the circular polarization is not found for 
larger $k$ modes (e.g., for ${\bf k}=(2,0,0)$). 
This is a very special property of the 
$k=1$ modes, that
they evolve in a quasi force-free state in time. This 
condition is consistent with the 
ensemble average predictions 
of the Gibbsian theory \cite{FrischEA75,StriblingMatthaeus90},
but is not imposed by it as an exact condition due to allowance for fluctuations
about the equilibrium expectation.
We will come back to this issue in the discussion section.

\subsubsection{Spherical geometry}

\begin{figure}
\epsfysize=17cm
\centerline{\epsffile{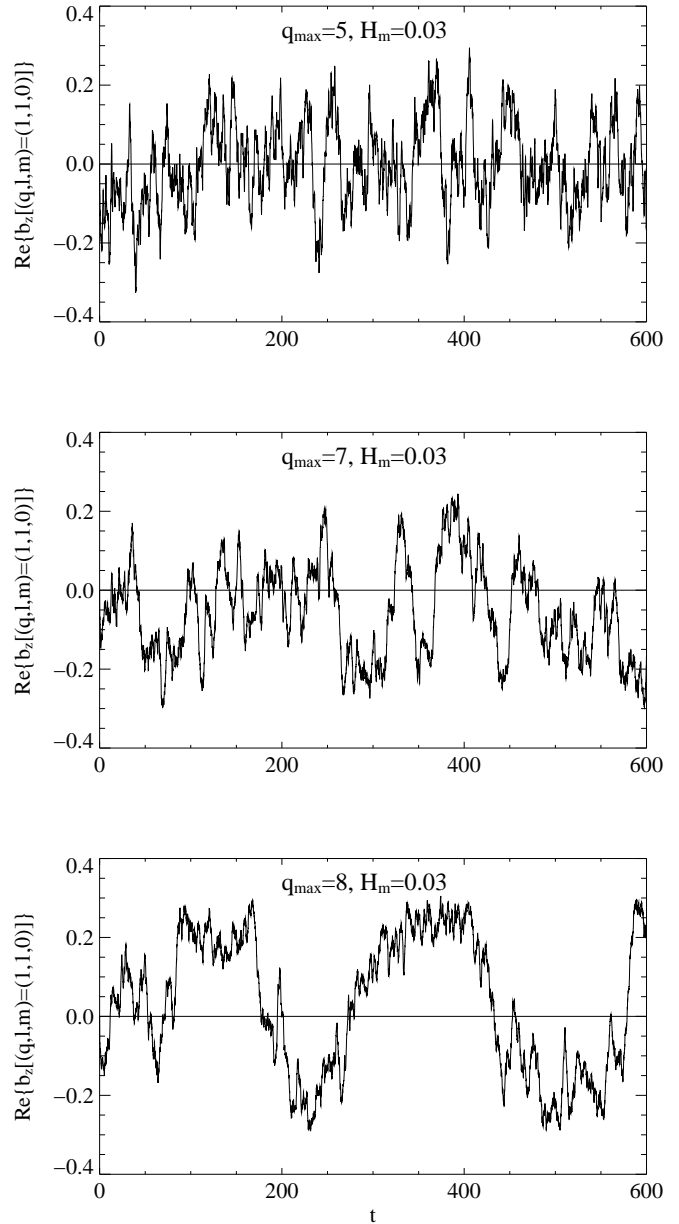}}
\caption{
Time series of the $b_z$ component of the magnetic field
for the $(q,l,m)=(1,1,0)$ mode in 
the spherical MHD run with $q_{max}=5, 7$, and $8$, and
magnetic helicity $H_m=0.03$.
}
\label{fig:timeseriesqm}
\end{figure}

\begin{figure}
\epsfysize=17cm
\centerline{\epsffile{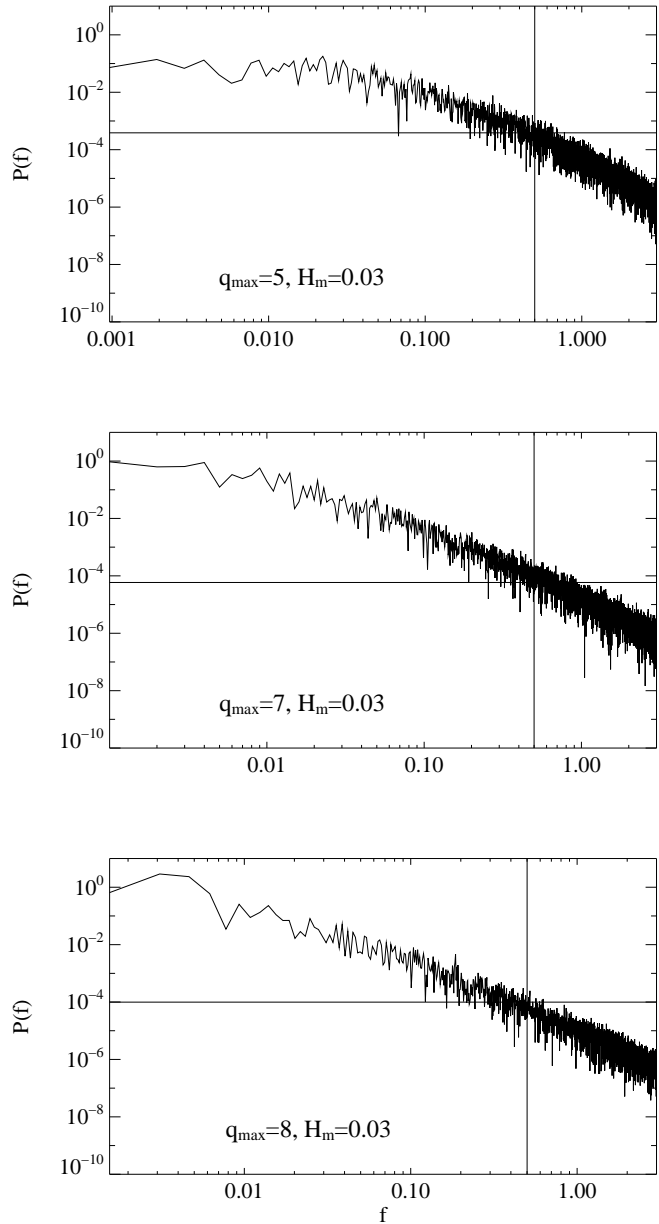}}
\caption{
Power frequency spectra of the time series 
in Figure (\ref{fig:timeseriesqm}), for $q_{max}=5, 7$, and $8$, and
$H_m=0.03$.
The reference vertical line corresponds to a frequency value $f_L=1/2$
(see text) and the horizontal line to the value of $P(f_L)$.
\label{fig:freqspectrqm}
}
\end{figure}

The next example corresponds to a series of spherical MHD runs. 
Figure (\ref{fig:timeseriesqm}) 
shows three runs with a fixed value of $H_m=0.03$,
and with three different values  of 
$q_{max}=5, 7$, and $8$, corresponding to increasing 
number of degrees of freedom in the model. 
Here, $q_{max}$ is the
maximum value of the $q$ index corresponding to the radial number
in a spherical harmonic expansion
[see Eqs. (\ref{eq:ji})-(\ref{eq:psi}) and text therein]; 
the maximum values of $l$ and $m$ 
are also increased accordingly.
In this particular geometry, 
the maximum possible value of helicity for a flow with unit 
energy is $\approx 0.22$.
 Again, as in 
the periodic box runs, it is seen that
increasing the size of the system shows the appearance of long
time fluctuations and a corresponding range of $1/f$ power frequency
spectrum, as shown in the plots of Figure (\ref{fig:freqspectrqm}).

Here we show again in Figure (\ref{fig:freqspectrqm}) references
vertical line for the frequency $f_L=1/2$ based on the largest
local non-linear time $T_L=L/U_L=2$ which corresponds to structures with
the diameter of the sphere $L=2$ (unit radius) and a unit root
mean square velocity $U_L=1$.

\begin{figure}
\epsfysize=12cm
\centerline{\epsffile{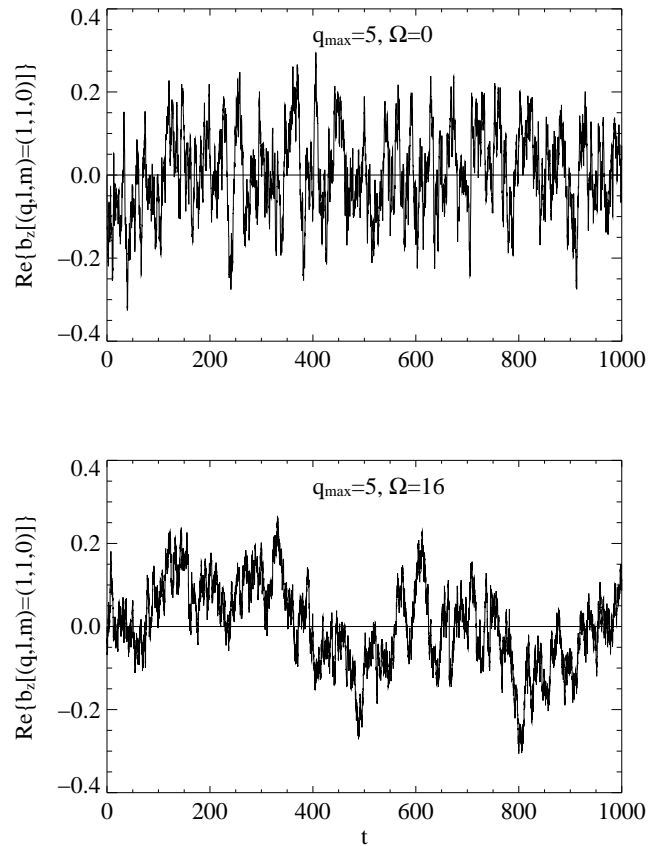}}
\caption{
Time series of the $b_z$ component of the magnetic field
for the $(q,l,m)=(1,1,0)$ mode in 
the spherical MHD run with zero rotation ($\Omega=0$, top), and $\Omega=16$ (bottom)
and $q_{max}=5$.
\label{fig:timeseriesOm}
}
\end{figure}

\begin{figure}
\epsfysize=12cm
\centerline{\epsffile{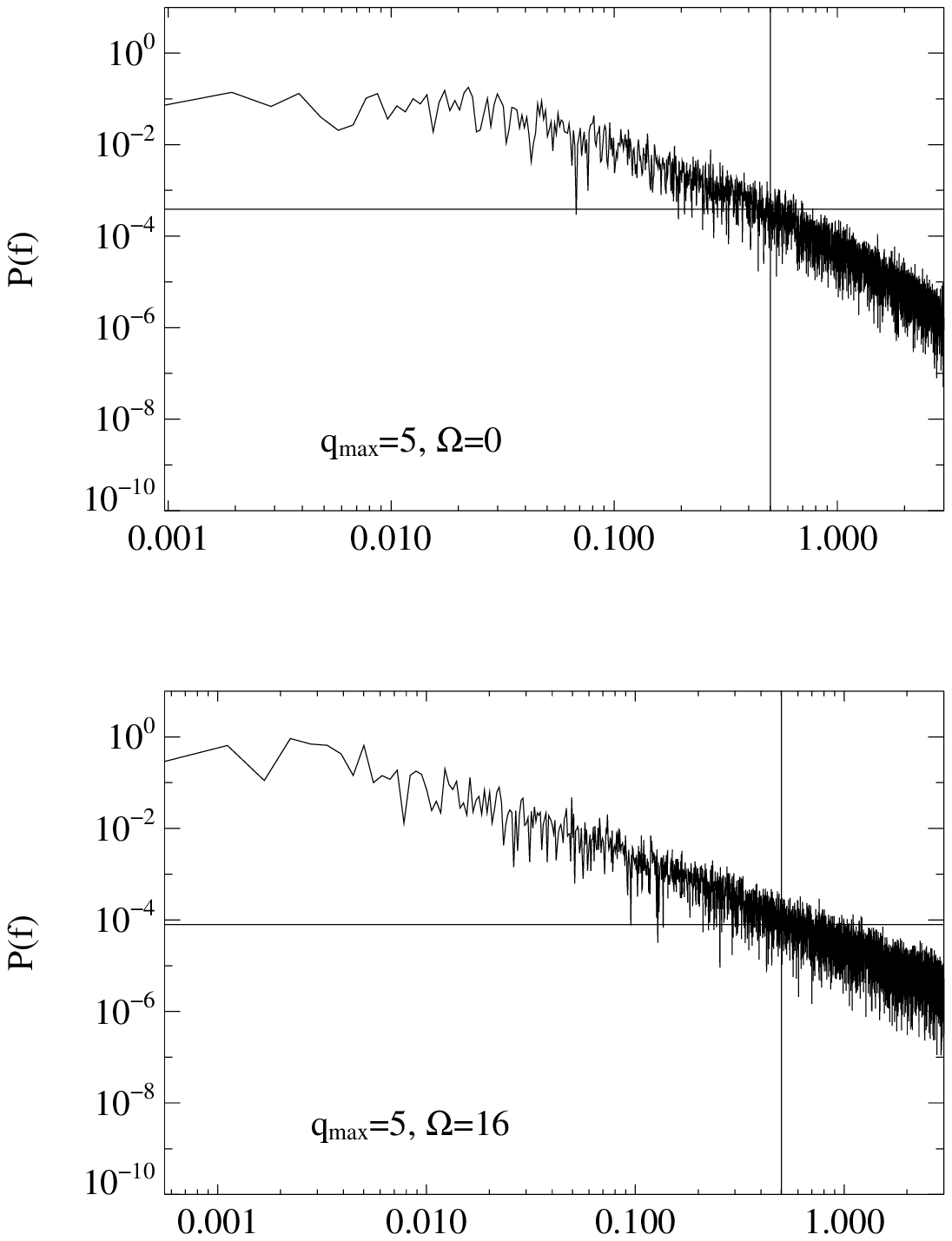}}
\caption{
Power frequency spectra of the time series 
in Figure (\ref{fig:timeseriesOm}), with rotation $\Omega=0$ and $16$ and
$q_{max}=5$.
The reference vertical line corresponds to a frequency value $f_L=1/2$
(see text) and the horizontal line to the value of $P(f_L)$.
\label{fig:freqspectrOm}
}
\end{figure}

Another effect is shown in Figure (\ref{fig:timeseriesOm}) which
corresponds to the time behavior of a single mode, by including
non-zero rotation, with $\Omega=16$. Results for size $q_{max}=5$ 
with $\Omega=0$ and $\Omega=16$ are
shown.
It is seen that longer time fluctuations 
appear with the addition of rotation in the system.
The corresponding frequency spectra in 
Figure (\ref{fig:freqspectrOm}) also show a wider range
of $1/f$ noise for the rotating case.

\begin{figure}
\epsfysize=9cm
\centerline{\epsffile{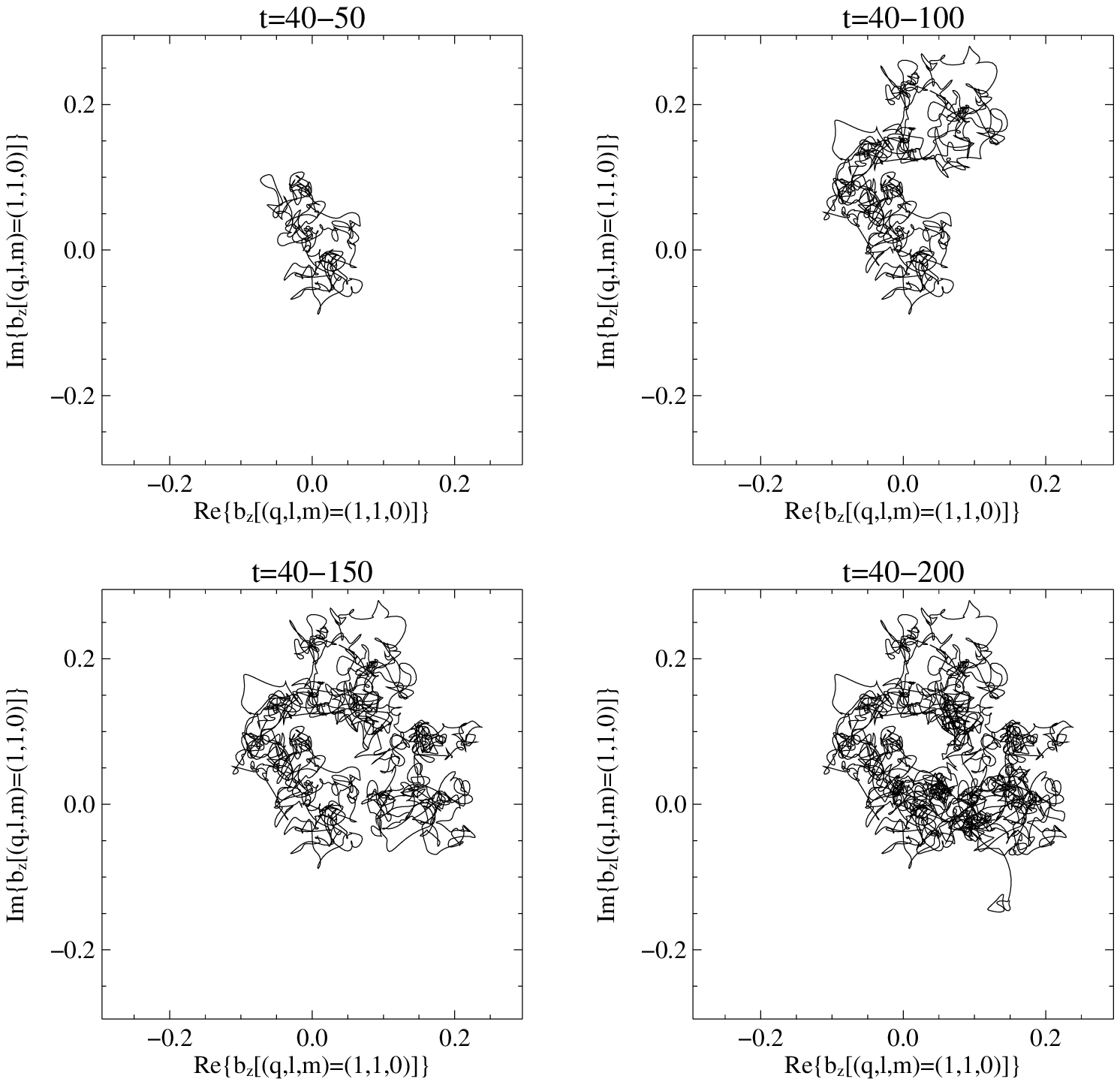}}
\caption{
Complex phase space trajectory for the $(q,l,m)=(1,1,0)$ mode in the 
spherical MHD run, with $q_{max}=5$ and rotation $\Omega=16$.
}
\label{fig:pspaceq}
\end{figure}

The long time fluctuations for the case of $q_{max}=5$ and $H_m=0.03$
with rotation (see Figure (\ref{fig:timeseriesOm})) appear 
as sign reversals of the $z$-component of the magnetic field,
with excursions with periods of the order $T \sim 50 - 200$.
In fact, another quantity
that can be studied for this system is the magnetic dipole, which
is essentially a weighted moment average of the magnetic field,
which tends to highlight the low $q$ modes. We defer a 
detailed analysis of the magnetic dipole reversals to another paper, 
but we point out that this long time fluctuation phenomenon is 
directly related to
$1/f$ noise and
delayed ergodicity of the $q=1$ mode in complex
phase space. As an example, a phase space plot of the temporal behavior of the $q=1$ mode
 is shown in Figure (\ref{fig:pspaceq}). It can be seen that this mode
spends a long time in a restricted region of phase space, thus delaying 
overall ergodicity.

\subsection{Three-dimensional MHD and RMHD with a background magnetic
field in a cubic box}

\subsubsection{MHD with a background magnetic field}

In the presence of a background uniform magnetic field ${\bf B_0}$, the
3D MHD equations lose one of the quadratic invariants, the
magnetic helicity, so the two invariants that remain are
the total energy (magnetic plus kinetic) and the cross helicity.
As will be discussed in a following section, this fact can have
an influence on the interaction of the lowest $k=1$ mode with
the remaining modes in the system, and correspondingly on the
$1/f$ power spectrum. 

\begin{figure}
\epsfysize=6cm
\centerline{\epsffile{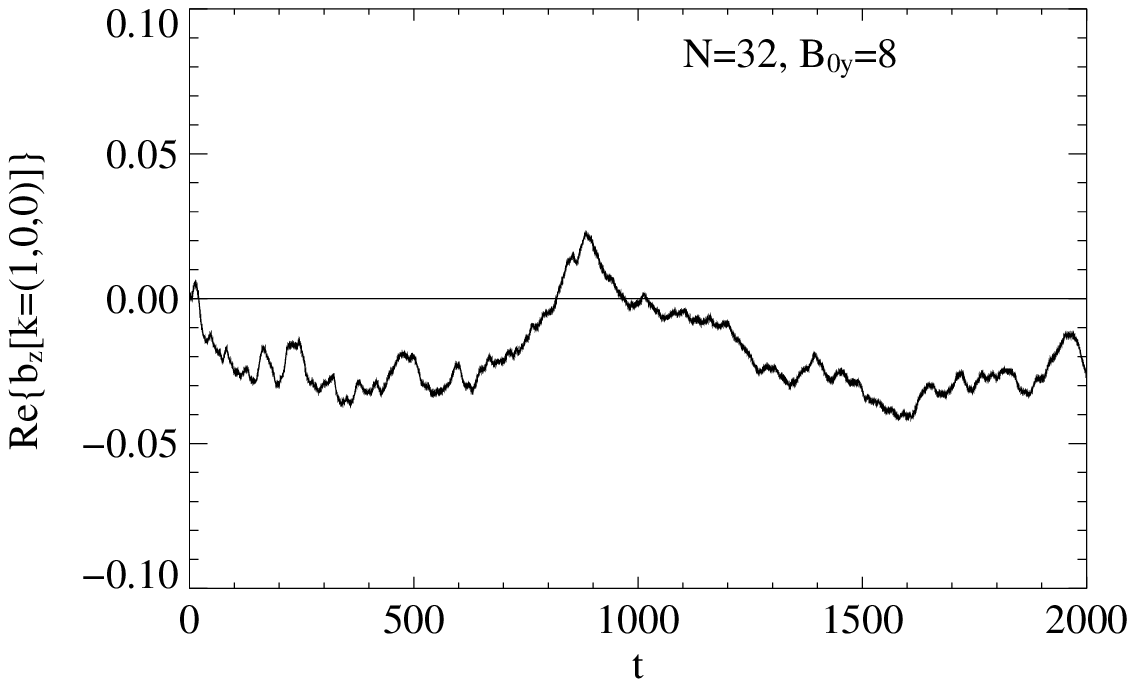}}
\caption{
Time series of the z-component of the magnetic field
for the ${\bf k}=(1,0,0)$ mode in 
the MHD $32^3$ run with background magnetic field $B_{0_y} = 8$ and $b_{rms}=1$.
}
\label{fig:timeB0}
\end{figure}

\begin{figure}
\epsfysize=6cm
\centerline{\epsffile{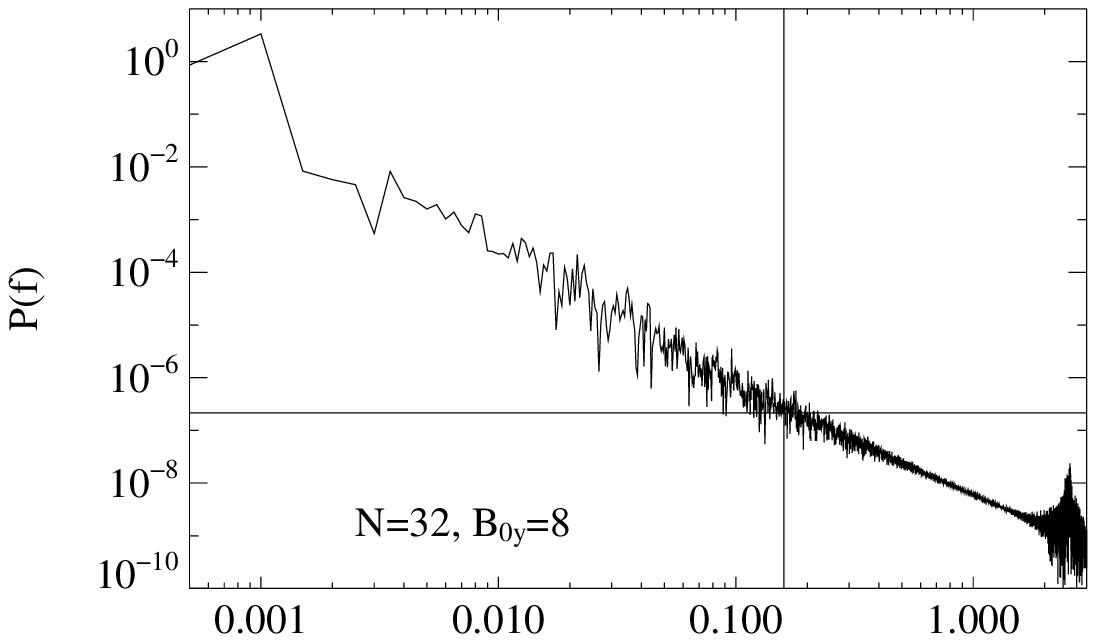}}
\caption{
Power frequency spectra of the time series shown in 
Figure (\ref{fig:timeB0}).
The reference vertical line corresponds to a frequency value $f_L=1/(2\pi)$
and the horizontal line to the value of $P(f_L)$.
}
\label{fig:freqspectrB0}
\end{figure}

\begin{figure}
\epsfysize=8cm
\centerline{\epsffile{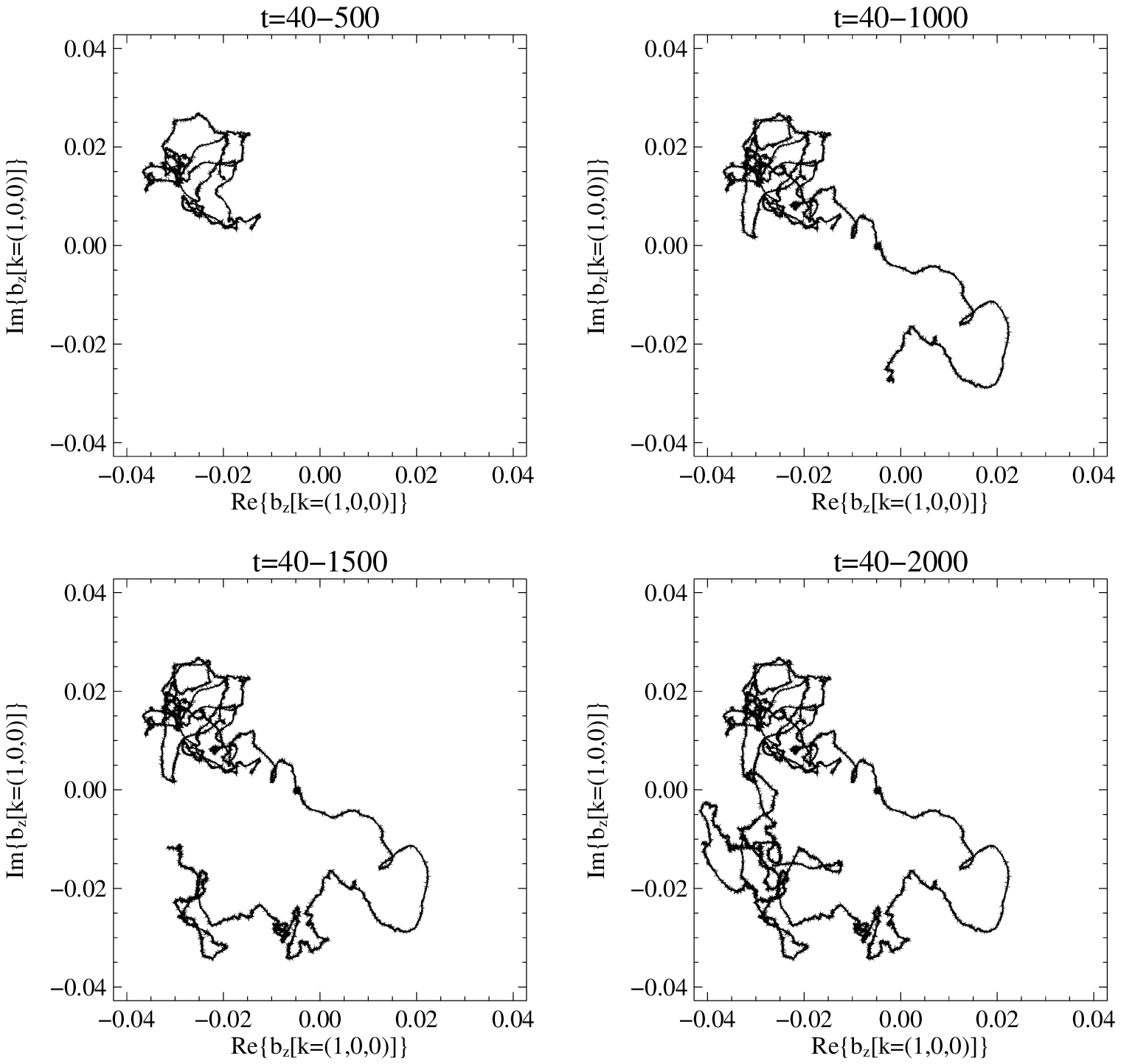}}
\caption{
Complex phase space trajectory for the ${\bf k}=(1,0,0)$ mode in the 
MHD $32^3$ run with background magnetic field $B_{0_y}=8$.
}
\label{fig:pspaceB0}
\end{figure}

Results for simulation runs of ideal 3D MHD with a background magnetic
field in the $y$-direction $B_{0_y}=8$ (as compared to fluctuations 
r.m.s.~values 
of $\left< b^2\right> ^{1/2}=b_{rms}=1$) 
are shown in Figure (\ref{fig:timeB0}), for the time evolution
of a fluctuating component of the magnetic field of the 
${\bf k}=(1,0,0)$ mode, and 
for the corresponding frequency spectra in Figure (\ref{fig:freqspectrB0}). 
This corresponds to a simulation of size $32^3$.
Long time fluctuations of the order of $1000$ unit
times appear, and a corresponding strong enhancement of very low frequency power, 
associated with a $1/f$ spectrum, 
is clearly observed. This case of MHD with a background magnetic
field has been thoroughly studied for the driven/dissipative case \cite{DmitrukMatthaeus07}. Here, it can
be seen that the $1/f$ spectrum appears even in the ideal case
and we conclude that it is 
a property of the non-linear interactions among
the modes in the system, and does not depend 
on the presence of dissipation in an essential way.
The complex plane phase space trajectory for 
this case is shown in Figure (\ref{fig:pspaceB0}), 
and delayed ergodicity of the 
${\bf k}=(1,0,0)$ mode is also clearly observed. 

\subsubsection{Reduced MHD}

\begin{figure}
\epsfysize=6cm
\centerline{\epsffile{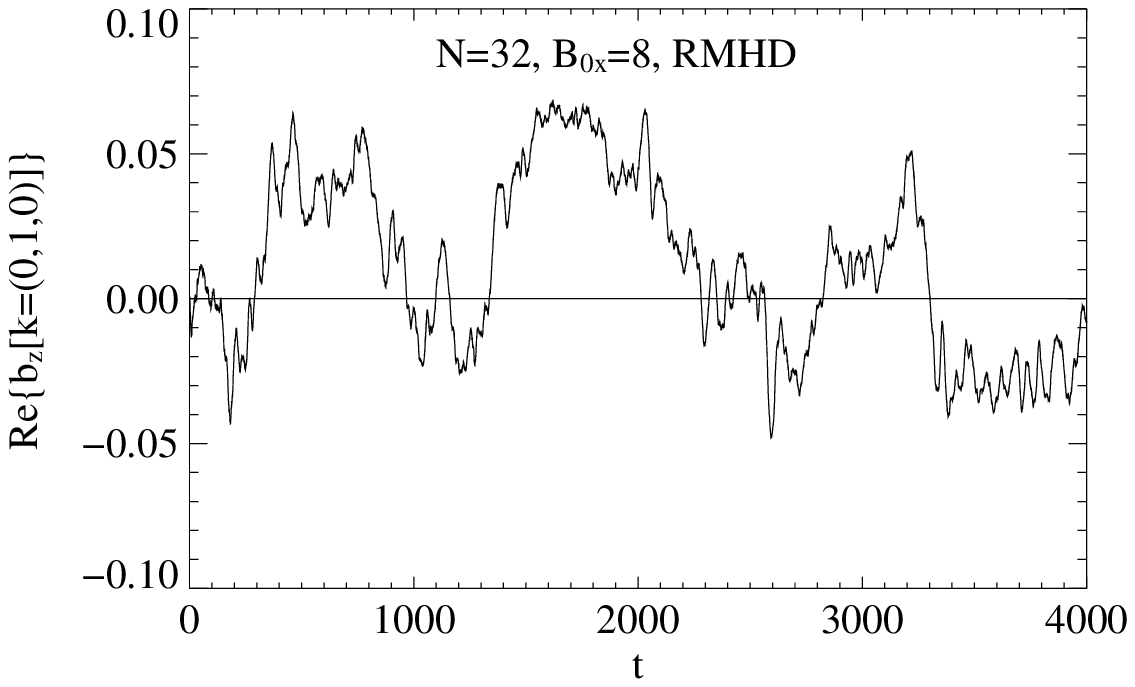}}
\caption{
Time series of the z-component of the magnetic field
for the ${\bf k}=(0,1,0)$ mode in 
the RMHD $32^3$ run with background magnetic field $B_{0_x} = 8$.
}
\label{fig:timermhd}
\end{figure}

\begin{figure}
\epsfysize=6cm
\centerline{\epsffile{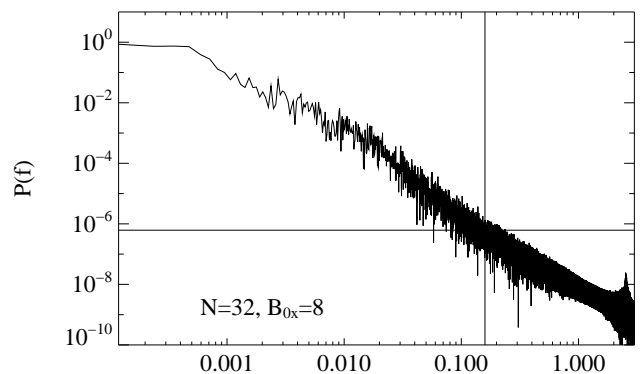}}
\caption{
Power frequency spectra of the time series shown in 
Figure (\ref{fig:timermhd}). 
The reference vertical line corresponds to a frequency value $f_L=1/(2\pi)$
and the horizontal line to the value of $P(f_L)$.
}
\label{fig:freqspectrrmhd}
\end{figure}

Another related case of interest is given by the RMHD system, which 
is, as discussed in Section II, an approximation to the 
low-frequency dynamics of the MHD equations 
with a large background magnetic field. 
Results for a RMHD case
are shown in Figure (\ref{fig:timermhd}), namely 
time series for a component of the 
fluctuating magnetic field for the ${\bf k}=(0,1,0)$ mode, 
and corresponding power frequency spectrum in 
Figure (\ref{fig:freqspectrrmhd}). This is for
a case with a background magnetic field in the $x$-direction
$B_{0_x}=8$ and size $32^3$.
As can be seen, long time fluctuations are evident, with
periods on the order of $1000$ unit times. 
The RMHD system admits 
two known quadratic ideal invariants, 
the total energy and the 
cross helicity.
It is interesting however that an additional quantity, 
the mean 
square vector potential,
which is strictly
invariant in ideal 2D MHD,
behaves as a quasi-invariant in RMHD \cite{ServidioCarbone05}. 
By this we mean that this quantity
remains statistically constant for long periods of time.
The connection of this with the emergence of long time fluctuations
will be further discussed in section IV.

\subsection{Three-dimensional hydrodynamic with and without rotation}

There are two quadratic invariants in ideal three-dimensional hydrodynamics,
the kinetic energy 
\begin{equation}
E_u= \frac{1}{2} \langle  |{\bf u}|^2 \rangle,
\end{equation}
and the kinetic helicity 
\begin{equation}
H_v= \langle {\bf \omega} \cdot {\bf u} \rangle \,  .
\end{equation}
These are also invariants of the 3D HD 
equations with addition of constant rotation.

In the absence of rotation, these invariants do not 
condense to the longest wavelength, or engage preferentially in back-transfer
to the lowest $k$ values 
\cite{Kraichnan73,KrstulovicEA09}.
Lacking these tendencies a rationale is lacking for expecting 
an inverse cascade to large scales in the dissipative case. 
In the 
presence of rotation, however, and in the ideal case, there is a transient 
reduction to quasi two-dimensional behavior and a transient 
condensate, associated with the quasi-conservation of the energy in 
two-dimensional modes (modes with $k_\parallel=0$, where parallel refers 
to the direction of the rotation axis) \cite{Bourouiba08,Mininni11DMP}. 
These modes correspond to the so-called slow manifold of the system. 
In the forced-dissipative rotating case, resonant interactions transfer 
energy toward these modes, also resulting in two-dimensionalization and 
in the development of an inverse energy cascade \cite{Cambon89,Waleffe93}. 

The ideal behavior discussed above is similar to the behavior of MHD 
in the presence of a background magnetic field (see previous 
subsection), which also reduces to a quasi two-dimensional dynamical 
behavior and has the mean square vector potential as a quasi-invariant 
\cite{Shebalin83,ServidioCarbone05}. As will be discussed further in the next section, 
this has an effect on the emergence of long time fluctuations (see also \cite{ServidioCarbone05}). 

\begin{figure}
\epsfysize=12cm
\centerline{\epsffile{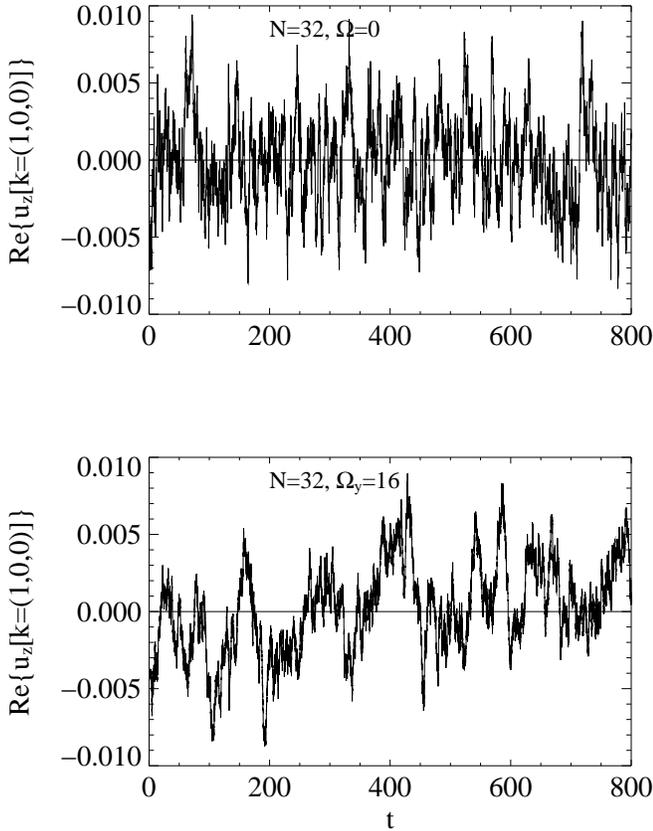}}
\caption{
Time series of the $z$-component of the velocity field
for the ${\bf k}=(1,0,0)$ mode in 
the HD $32^3$ run with rotation $\Omega=0$ (top) and $\Omega_y= 16$ (bottom).
}
\label{fig:timehd}
\end{figure}

\begin{figure}
\epsfysize=12cm
\centerline{\epsffile{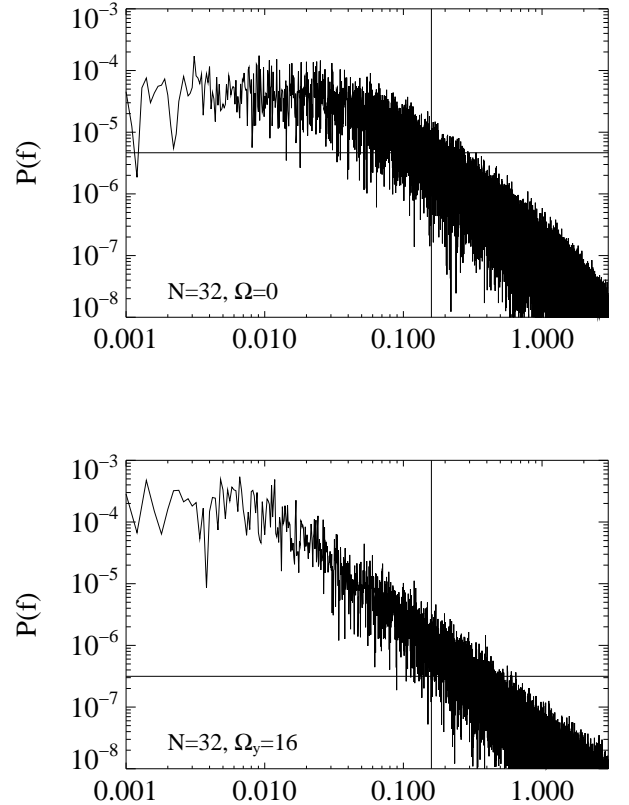}}
\caption{
Power frequency spectra of the time series of 
Figure (\ref{fig:timehd}). 
The reference vertical line corresponds to a frequency value $f_L=1/(2\pi)$
and the horizontal line to the value of $P(f_L)$.
}
\label{fig:freqspectrhd}
\end{figure}

The time evolution of a velocity component for the ${\bf k}=(1,0,0)$ 
mode with no rotation and with rotation $\Omega_z=16$ (along the $z$-axis) 
in a periodic box is shown in Figure (\ref{fig:timehd}), 
and the corresponding
frequency spectra are shown in Figure (\ref{fig:freqspectrhd}).
Long time
fluctuations are much more apparent in the case with rotation, 
where fluctuations are observed 
with periods of the order of 
$T \sim 50-100$, together with a corresponding 
enhancements in the 
low frequency
$1/f$ range of the spectrum. 
Similar results are obtained for simulations in 
spherical geometry
(not shown).

\section{Discussion}

All the simulations discussed in the previous section are 
summarized 
in Table I. We summarize now the results and discuss the 
related phenomena of
emergence 
of long time fluctuations, delayed ergodicity, and corresponding $1/f$ 
power law in the frequency spectra.

\begin{table*}
\caption{
Summary of all runs,
with $N^3$ the total number of grid points, $H_m$ and $H_v$ the magnetic and kinetic helicity, and $\Omega$ or $B_0$ the respective amplitude of the imposed solid body rotation or uniform magnetic field. The first column indicates what equations have been integrated, SMHD being the MHD equations in spherical geometry with $q_{max}$ corresponding to the truncation of the C-K functions (see equation (5) and text);  all other runs are computed in periodic cubic geometry. ``ROT" stands for runs with rotation, and $B_0$ stands for runs with an imposed magnetic field. The last column gives the estimated strength and range of the observed $1/f$ noise spectrum (see Figures).
}
\begin{ruledtabular}
\begin{tabular}{ccccc}
Run & $N^3$ & $H_m$ or $H_v$ & $\Omega$ or $B_0$ & $1/f$ range \\
\hline
MHD & $16^3$ & $-0.008$ & $0$ & weak, $T \sim 20-100$ \\
MHD & $16^3$ & $0.027$ & $0$ & medium, $T \sim 20-200$  \\
MHD & $16^3$ & $0.129$ & $0$ & strong, $T \sim 100-1000$ \\
MHD & $16^3$ & $-0.395$ & $0$ & strong, $T \sim 200-1000$ \\
MHD & $32^3$ & $0.027$ & $0$ & strong, $T \sim 100-1000$ \\
MHD & $64^3$ & $0.015$ & $0$ & strong, $T \sim 1000-2000$  \\
SMHD & $q_{max}=5$ & $0.03$ & $0$ & weak, $T \sim 10-50$ \\
SMHD & $q_{max}=7$ & $0.03$ & $0$ & strong, $T \sim 50-100$ \\
SMHD & $q_{max}=8$ & $0.03$ & $0$ & strong, $T \sim 100-200$ \\
SMHD + ROT & $q_{max}=5$ & $0.03$ & $16$ & strong, $T \sim 50-200$ \\
MHD + $B_0$ & $32^3$ & - & $8$ & strong, $T \sim 500-1000$ \\
RMHD & $32^3$ & - & $8$ & strong, $T \sim 500-1000$ \\
HD & $32^3$ & $-0.26$ & $0$ & weak, $T \sim 10-50$ \\
HD + ROT & $32^3$ & $-0.26$ & $16$ & medium, $T \sim 50-100$ \\
\end{tabular} \end{ruledtabular} \end{table*}

The first examined case corresponds to ideal 3D MHD. This system shows 
long term memory, $1/f$ noise, and delayed ergodicity in the $k=1$ 
modes. As pointed out, the system has three quadratic invariants, the 
total energy, cross helicity, and magnetic helicity, and in a 
statistical steady state the amplitudes of Fourier modes are 
controlled by the Gibbs ensemble prediction. Specifically, the magnetic
helicity allows condensation at the lowest wavenumber mode. This
happens for non-zero values of $H_m$, and it becomes more intense
as the absolute value of $H_m$ is increased. Condensation
also become more intense as the number
of modes $\sim N^3$ is increased for fixed $H_m/E$. 
The $k=1$ mode is special
then in the dynamics of this system. As shown, this mode is in 
a quasi force-free 
state, so its evolution is slow, weakly coupled
with a sea of lower amplitude modes at much larger wavenumbers. The 
force-free property is prescribed by the Gibbs ensemble solution;
the modes with $k=1$ have maximum helicity,
the 
magnetic field is parallel to the current density, and these
largest scale fluctuations
are circularly polarized.

The coupling between the $k=1$ modes and the small scale modes is 
defined by triads of wavenumbers, constructed
with a $k=1$ mode and two large wavenumber modes.
The time evolution of these interactions is controlled by
the (comparatively small) amplitude of the larger wavenumber modes,
but the large length-scale (small $k$) of the lowest wavenumber mode.
As pointed out in \cite{DmitrukMatthaeus07},
this can be seen from the expression for this type of interaction
which is of the (schematic) form
\begin{equation}
\frac{\partial b(k)}{\partial t} =
-i k \sum_{k=p+q} u(q) b(p) \ ,
\end{equation}
where $b(k), u(q), b(p)$ are generic Fourier mode amplitudes,
with the constraint that ${\bf k}={\bf p}+{\bf q}$. In particular,
we consider the lowest wavenumber mode $k=1$. If the triadic interaction
is local, then $k \sim p \sim q$ by definition and the timescale of that interaction is
given by $\lbrack k u(k=1) \rbrack^{-1} \sim 1$,
whereas if the interaction is nonlocal,
then $p,q \gg k=1$, $p \sim q$ and
the timescale is $\lbrack k u(q) b(q)/b(k=1) \rbrack^{-1}$
which is much longer than the local
timescale since $u(q), b(q) \ll u(k=1), b(k=1)$.

The fact that $1/f$ noise is also observed in cases with low magnetic 
helicity but increasing size $N$ suggests however that even 
in cases with amplitude equipartition among modes (i.e., when 
magnetic helicity is small, as indicated by the Gibbs 
ensemble predictions), the sea of large wavenumber modes
is again slowly modifying the dynamics of the $k=1$ mode. This idea
is supported by previous studies \cite{KrstulovicEA09} in which 
the effect of the large wavenumber modes on the lower wavenumber modes
is modeled through an effective viscosity, even though the systems
are strictly ideal, like the Euler equations. This is also similar
to ideas suggested by low dimensional dynamical systems 
in connection with the reversals of the geomagnetic field, see e.g. 
\cite{Benzi05,Sorriso-ValvoEA07}, with bi-stable
states driven by noise (in this case, the large wavenumber modes 
would act as the driving noise for large-scale behavior).

These results, besides being obtained for different resolutions and 
values of the invariants in ideal cases (i.e., without viscous or 
external forcing, indicating the long term behavior is intrinsic 
to the system of equations), are also obtained for two different 
geometries: in periodic boxes and in spheres. As a result, we conclude that the boundary 
conditions do not seem to affect the long term behavior.

As mentioned before, the 3D MHD system has an ideal invariant that 
experiences 
a condensation to the
longest wavelength modes of the system, 
and in the dissipative driven case is expected to 
be involved in an inverse cascade. 
Interestingly, another set of results observed here suggest that certain systems
like MHD with a background magnetic field or HD with rotation also
allow for the emergence of long time fluctuations and $1/f$ behavior. These
systems do not have an ideal invariant 
that condenses to the lowest
wavenumber mode in the Gibbs ensemble.
There is nevertheless a slow manifold of modes
(the two-dimensional modes) 
that is distinguished from all other degrees of freedom. We argue 
that this slow manifold is controlling the emergence of long time 
fluctuations in the $k=1$ modes, as observed in the results. It is 
interesting to note that these systems are characterized by the 
existence of a quasi-invariant. By this 
we mean 
a slowly varying but 
not strictly conserved
quantity, such as 
the square vector potential 
$\left<a^2\right>$ in MHD with a background field, or the energy in 
modes with $k_\parallel=0$ in HD with rotation. 
The quasi-invariants
are approximately conserved in the slow 
manifold, 
and thus introduce longer timescales in the dynamics
by permitting transient condensation at 
the lowest wavenumber mode.

Finally, a third category is that of systems with flat frequency 
spectrum for small frequencies, which are associated with 
negligible long time correlations. An example is 
3D NS without rotation, or 3D MHD with no magnetic helicity and zero mean magnetic field. 
These systems have no 
invariant or quasi-invariant that condensates at large scales in 
the ideal case.

\section{Conclusion}

The emergence of long time fluctuations and $1/f$ noise in the frequency
spectrum of field variables is observed in systems with a quadratic invariant allowing
condensation at the lowest wavenumber mode, like the magnetic 
helicity in 3D MHD. This happens when the invariant is large but also
when it is small, provided the number of modes ($N^3$) is large
enough. This happens indistinctly in geometries like a periodic box
or a sphere, thus we argue that this is an intrinsic property of
the non-linear couplings in the system and is not dependent on the
geometry, or other external properties of the system like driving,
or dissipation, which are absent in the ideal systems analyzed here.

We can conjecture therefore that long time fluctuations will be also 
observed in ideal 2D MHD and 2D HD, where a quadratic invariant 
allowing condensation at the lowest wavenumber mode also exists. 
A previous indication of this is given in \cite{DmitrukMatthaeus07} 
where the driven-dissipative case for these systems
is studied and showed to have $1/f$ 
noise. Other non-linear systems that do not belong to fluid dynamics, 
but have Gibbsian statistical condensates and 
dissipative-driven inverse cascades, may also show this behavior 
(see for instance a case in quantum optics \cite{LvovNewell00}).

The $1/f$ power spectrum is also observed in systems with a
slow manifold and quasi-invariants, like MHD with a background magnetic 
field and hydrodynamics with rotation. We also argue that other 
systems, like flows in the geostrophic approximation, with a slow 
manifold dynamics, may show long time fluctuations as well. 

The observed occurrence of 
a range of $1/f$ frequency spectra in 
interplanetary magnetic field and density fluctuations
is another case of relevance to the present discussions 
\cite{MatthaeusGoldstein86,MatthaeusEA07,BemporadEA08}
This signal is observed in the frequency 
range $\sim 10^{-5} - \sim 10^{-4}$ Hz
in the solar wind at 1AU and beyond, 
but is also observed in the coronal and in the solar photospheric 
magnetic field. It is therefore possible that the origin 
of this signal is either in the solar dynamo or in coronal 
dynamics, or both. 
Interestingly one might well expect slow manifold behavior in
either of these cases, due to rotation or
regional effects of magnetic helicity in the dynamo, 
or quasi-invariance of the mean square potential in the corona, which is 
dominated by a strong large scale magnetic field.
It is tempting to also associate this phenomenon to 
long-time memory effects in geophysical 
flows, as e.g., the ocean circulation; indeed, the origin of multi-decadal time-scales 
in the climate evolution still remains mysterious since it is not directly associated 
with a known instability and yet it is well observed, and reproduced, with approximate accuracy, 
in numerical models (see e.g. \cite{c1,c2,c3,c4}).

Finally, three-dimensional hydrodynamic flows, which do not have 
condensed invariants or quasi-invariants, do not develop $1/f$ noise 
in the ideal case (see also \cite{DmitrukMatthaeus07} for similar 
results in the forced-dissipative case). The development of 
$1/f$ noise in such systems, if it happens, may be associated with 
forcing or boundary conditions.

The main result that can be concluded here is that in many cases,
long-time fluctuations are intrinsically given by the non-linear 
dynamics of the system, and not controlled by external properties or dissipation. 
Furthermore these studies serve to 
strengthen the growing understanding
that there are deep connections
between condensation (or quasi-invariants), 
and $1/f$ signals at low frequency, as well 
between the associated slow manifolds and 
irregular large scale stochastic reversals (or delayed ergodicity).  
A number of phenomena observed in many types of flows can be all 
understood in terms of these couplings, and may therefore have 
universal properties linked to the presence of invariants.

\begin{acknowledgments}
Computer time was provided by the University of Delaware, 
the University of Buenos Aires and NCAR.
NCAR is sponsored by the National Science Foundation. PD and 
PDM acknowledge support from grants UBACYT 20020090200602 and 20020090200692, 
PICT 2007-02211 and 2007-00856, and PIP 11220090100825. AP and PDM 
acknowledge support from NSF-CMG grant AGS-1025183. WHM acknowledges 
support from NSF grant ATM-0539995 and ATM-0752135 (SHINE) and NASA
Heliospheric Theory Program grant NNX08AI47G.

\end{acknowledgments}

\end{document}